\documentstyle{mn}

\newif\ifAMStwofonts
\AMStwofontstrue

\ifoldfss
  \ifCUPmtlplainloaded \else
    \NewTextAlphabet{textbfit} {cmbxti10} {}
    \NewTextAlphabet{textbfss} {cmssbx10} {}
    \NewMathAlphabet{mathbfit} {cmbxti10} {} 
    \NewMathAlphabet{mathbfss} {cmssbx10} {} 
  \fi
  \ifAMStwofonts
    \ifCUPmtlplainloaded \else
      \NewSymbolFont{upmath} {eurm10}
      \NewSymbolFont{AMSa} {msam10}
      \NewMathSymbol{\upi}     {0}{upmath}{19}
      \NewMathSymbol{\umu}     {0}{upmath}{16}
      \NewMathSymbol{\upartial}{0}{upmath}{40}
      \NewMathSymbol{\leqslant}{3}{AMSa}{36}
      \NewMathSymbol{\geqslant}{3}{AMSa}{3E}

       \let\le=\leqslant
       \let\ge=\geqslant
    \fi
  \fi
\fi 

\ifnfssone
  \newmathalphabet{\mathit}
  \addtoversion{normal}{\mathit}{cmr}{m}{it}
  \addtoversion{bold}{\mathit}{cmr}{bx}{it}
  \newmathalphabet{\mathbfit} 
  \addtoversion{normal}{\mathbfit}{cmr}{bx}{it}
  \addtoversion{bold}{\mathbfit}{cmr}{bx}{it}
  \newmathalphabet{\mathbfss} 
  \addtoversion{normal}{\mathbfss}{cmss}{bx}{n}
  \addtoversion{bold}{\mathbfss}{cmss}{bx}{n}
  \ifAMStwofonts
    \ifCUPmtlplainloaded \else
      \UseAMStwoboldmath
      \makeatletter
      \new@mathgroup\upmath@group
      \define@mathgroup\mv@normal\upmath@group{eur}{m}{n}
      \define@mathgroup\mv@bold\upmath@group{eur}{b}{n}
      \edef\UPM{\hexnumber\upmath@group}
      \new@mathgroup\amsa@group
      \define@mathgroup\mv@normal\amsa@group{msa}{m}{n}
      \define@mathgroup\mv@bold\amsa@group{msa}{m}{n}
      \edef\AMSa{\hexnumber\amsa@group}
      \makeatother
      \mathchardef\upi="0\UPM19
      \mathchardef\umu="0\UPM16
      \mathchardef\upartial="0\UPM40
      \mathchardef\leqslant="3\AMSa36
      \mathchardef\geqslant="3\AMSa3E

       \let\le=\leqslant
       \let\ge=\geqslant
    \fi
  \fi
\fi 

\ifnfsstwo
  \DeclareMathAlphabet{\mathbfit}{OT1}{cmr}{bx}{it}
  \SetMathAlphabet\mathbfit{bold}{OT1}{cmr}{bx}{it}
  \DeclareMathAlphabet{\mathbfss}{OT1}{cmss}{bx}{n}
  \SetMathAlphabet\mathbfss{bold}{OT1}{cmss}{bx}{n}
  \ifAMStwofonts
    \ifCUPmtlplainloaded \else
      \DeclareSymbolFont{UPM}{U}{eur}{m}{n}
      \SetSymbolFont{UPM}{bold}{U}{eur}{b}{n}
      \DeclareSymbolFont{AMSa}{U}{msa}{m}{n}
      \DeclareMathSymbol{\upi}{0}{UPM}{"19}
      \DeclareMathSymbol{\umu}{0}{UPM}{"16}
      \DeclareMathSymbol{\upartial}{0}{UPM}{"40}
      \DeclareMathSymbol{\leqslant}{3}{AMSa}{"36}
      \DeclareMathSymbol{\geqslant}{3}{AMSa}{"3E}

       \let\le=\leqslant
       \let\ge=\geqslant
    \fi
  \fi
\fi 

\ifCUPmtlplainloaded \else
  \ifAMStwofonts \else 
    \def\upi{\pi}
    \def\umu{\mu}
    \def\upartial{\partial}
  \fi
\fi

\title[ARCS, The Arcminute Radio Cluster-lens Search --- I.]{ARCS, The Arcminute Radio Cluster-lens Search \\
	I. Selection Criteria and Initial Results}
\author[P.~M.~Phillips et al.]{P.~M.~Phillips,$^1$ I.~W.~A.~Browne,$^1$ P.~N.~Wilkinson$^1$ \\
$^1$University of Manchester, Jodrell Bank Observatory, Macclesfield, Cheshire, SK11 9DL}
\date{}

\pagerange{\pageref{firstpage}--\pageref{lastpage}}
\pubyear{2000}

\begin{document}

\maketitle

\label{firstpage}

\begin{abstract}
We present the results of an unbiased radio search for gravitational lensing events with image separations between 15\arcsec~and 60\arcsec, which would be associated with clusters of galaxies with masses $>10^{13~-~14}M_{\odot}$. A parent population of 1023 extended radio sources stronger than 35~mJy with stellar optical identifications was selected using the FIRST radio catalogue at 1.4~GHz and the APM optical catalogue. The FIRST catalogue was then searched for companions to the parent sources stronger than 7~mJy and with separation in the range 15\arcsec to 60\arcsec. Higher resolution observations of the resulting 38 lens candidates were made with the VLA at 1.4~GHz and 5~GHz, and with MERLIN at 5~GHz in order to test the lens hypothesis in each case. None of our targets was found to be a gravitational lens system. These results provide the best current constraint on the lensing rate for this angular scale, but improved calculations of lensing rates from realistic simulations of the clustering of matter on the relevant scales are required before cosmologically significant constraints can be derived from this null result. We now have an efficient, tested observational strategy with which it will be possible to make an order-of-magnitude larger unbiased search in the near future.
\end{abstract}


\section{Introduction}
Observations of gravitational lensing provide a probe of the distribution of mass in the universe on scales ranging from micro-lensing by sub-solar mass objects in the Galactic halo to weak-lensing by large-scale structures of the universe. In order to survey matter both light {\it and} dark on different size scales, unbiased systematic searches for lensing events within a well-defined sample of background sources must be carried out. Since the discovery of giant arcs in the region of clusters A370 and Cl2244 (Lynds \& Petrosian 1986; Soucail et al. 1987) more than 50 clusters have been found to contain arc-like images of background galaxies, including a clear example of multiple imaging by the cluster 0024+1654 (Colley, Tyson \& Turner 1996). The lensed images provide diagnostics for the cluster potential, allowing the distribution of mass to be mapped independently of the distribution of light. However, the currently known examples cannot tell us about the {\it general prevalence} of mass structures on the scale of clusters of galaxies (CGs) because they are in no sense statistically complete, having largely been discovered through observations of previously known clusters. Only by carefully examining a large well-defined sample of background sources for cases of gravitational lensing can we hope to place constraints on the general distribution of cluster-sized masses.

$N$-body cold-dark-matter (CDM) simulations of the universe make predictions about the degree of clustering of dark matter haloes at different epochs (for example Jenkins et al. 1998). The statistics of strong gravitational lensing events within a sample of background sources should therefore allow us to distinguish between models with different rates of cluster evolution. Early simulations by Wambsganss et al (1995) suggested that there could be as many strongly-lensed sources with image separations $>$5\arcsec, associated with groups and clusters of galaxies, as there are with image separations $<$5\arcsec~associated with single galaxy lenses. The JVAS and CLASS lens surveys (King et al. 1999; Myers 1996) found a lensing rate of approximately 1 in 500 for lens systems with image separations between 0\farcs3 and 6\arcsec. This suggests that in a search for lens systems associated with clusters of galaxies, a background source population of at least $\sim10^3$ is necessary before useful cosmological constraints can be set.

Previous searches for strong lensing events with image separations greater than a few arcseconds have started with quasar surveys because they provide a sample of objects with high mean redshift, which increases the lensing probability, and relatively low surface density, which reduces the number of false positives from random associations. Maoz et al. (1997) used a subsample of 76 quasars from the HST snapshot survey (Maoz et al. 1992) to look for lensing events with angular separations of 7\arcsec~to 50\arcsec. The fact that no lens systems were found allowed some extreme cosmological scenarios to be ruled out, for example that there exists a large population of `failed' or `dark' clusters. However, the small size of the parent sample limited the ability to place useful constraints on more conventional cosmological models.

In designing surveys to look for cluster-mass lenses, a basic problem arises when trying to recognise strongly-lensed images. As the image separation increases, the time delay between the images increases, and at a separation of tens of arcsec the characteristic delay can be hundreds of years. Basic properties of quasars such as the continuum emission, the emission line spectrum and the radio source nuclear structure may change significantly over this time. Due to their intrinsic variability, quasars are therefore not optimally-suited for use as probes for lensing by clusters of galaxies. A search method using background sources which are expected, a priori, to show no significant changes in physical properties over timescales of $\ga$1000~years is required to overcome this problem.

This paper is concerned with the Arcminute Radio Cluster-lens Search (ARCS), a project aimed at looking for strong gravitational lensing by galaxy clusters in a systematic and uniform manner. We use the fact that extended emission ($>$few~kpc) from an extragalactic radio source is unlikely to vary over the timescales of interest. In order to maximise our parent sample of background sources, whilst minimising the observational effort in following up potential lens candidates, we started from the existing FIRST 1.4~GHz survey (Becker, White \& Helfand 1995) and the Automatic Plate Measuring (APM) optical catalogue (for example Irwin, Maddox \& McMahon 1994). The FIRST survey provides both high sensitivity and sufficient resolution to allow us to reject most multiple sources as candidate lens systems directly before any follow-up observations have taken place. Furthermore, demanding stellar optical identifications of the radio sources from the APM catalogue both increases the mean redshift of the parent sample and provides a filter against selecting intrinsic double sources as lens candidates.

We describe our selection algorithm for identifying gravitational lens candidates from the FIRST and APM catalogues in section \ref{selection}. We then go on to describe the subsequent follow-up observations in section \ref{radio_obs}, including an analysis of the remaining CG-lens candidates at each observational stage. A review of the project and future work is laid out in section \ref{discussion}.

\section{Lens Candidate Selection}
\label{selection}
Our aim is to maximise the probability of detecting gravitational multiple images produced by CGs while minimising the observational effort required to confirm or refute them. As we pointed out above, reliable searches for lensing events on these scales require us to select background parent sources which should not vary over the relevant timescales which could extend to hundreds of years. We must also maximise the redshift of these background sources which increases the lensing probability. The FIRST survey combined with the APM catalogue derived from the Palomar Observatory Sky Survey (POSS) plates enables us to achieve these goals. 

\begin{figure}
\begin{center}
\setlength{\unitlength}{1cm}
\begin{picture}(5,3.5)
\put(-3.5,-6.5){\includegraphics{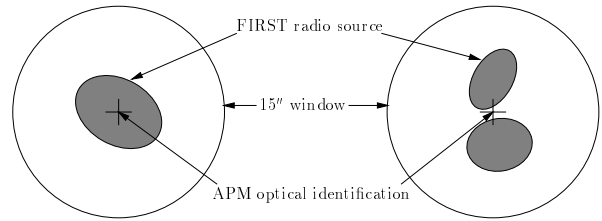}}
\end{picture}
\caption{Illustration of a one-component (left) and two-component (right) primary source. Note that the position of the optical identification for a two-component source is determined by the relative brightness of each component (see text). A range of two-component sources are acceptable (see Figure \ref{composite_source}) ranging from those with optical identification close to a relatively bright component, to those with two components of similar brightness and optical identification equidistant from each component.}\label{primary_source}
\end{center}
\end{figure}
\begin{figure}
\begin{center}
\setlength{\unitlength}{1cm}
\begin{picture}(4,4)
\put(-4,-7){\includegraphics{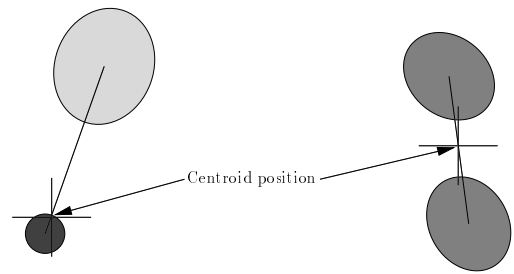}}
\end{picture}
\caption{Illustration of two extreme examples of two-component primary sources. The left-hand source has a centroid position near to a compact but relatively strong component whilst the second component is relatively extended but with significantly lower integrated flux density. This type will most likely be a `core-jet' source. The source on the right comprises two components with the same strength, and the centroid position is approximately equidistant from each component. In this case the source is probably a `classical double'.}\label{composite_source}
\end{center}
\end{figure}

\subsection{The FIRST Survey}
\label{FIRST}
The Faint Images of the Radio Sky at Twenty centimetres (FIRST) survey will eventually cover 10\,000 square degrees of the north galactic cap. The survey area will coincide with the Sloan Digital Sky Survey (SDSS), upon completion of which $\sim$50~per~cent of all FIRST sources will be optically identified to the $m(v)\sim24$ limit of SDSS. For our purposes, the north Galactic cap portion of the catalogue dated 98feb04 was used. This contains 382\,892 sources covering 4150 square degrees of the sky in the region RA$\simeq7^{\mathrm{h}}20^{\mathrm{m}}$ to RA$\simeq17^{\mathrm{h}}20^{\mathrm{m}}$, and $\delta\simeq$22\degr12\arcmin to $\delta\simeq$57\degr36\arcmin. The lower flux density limit is 1~mJy with a typical rms noise of 0.15~mJy and a beam-size of 5\farcs4; source positions are tied to the JVAS calibrator reference frame (Patnaik et al. 1992; Browne et al. 1998; Wilkinson et al. 1998) with a systematic uncertainty of $<$0\farcs03, while the overall systematic positional errors are $<$0\farcs1 throughout the survey area. Source parameters were derived by fitting elliptical Gaussian models. A more complete description of the FIRST catalogue can be found at URL {\tt http://sundog.stsci.edu/}.

Approximately 2.5~per~cent of `sources' in the FIRST catalogue are flagged as possible side-lobes of strong sources. Only about 15~per~cent of these sources are likely to be real, equivalent to less than 0.5~per~cent of the total. We therefore removed all possible side-lobes from our version of the catalogue prior to our analysis.

\subsection{The APM Catalogue}
\label{APM}
The APM catalogue is derived from the UK Schmidt and POSS plates. For our present purposes we used APM data obtained from the POSS for the region covered by FIRST. The APM catalogue contains data from blue O plates and red E plates to limiting magnitudes of 21.5 and 20 respectively, with a positional accuracy of $\sim$1\arcsec. Each detection is parameterised in terms of position, magnitude, and a classification into a stellar or a non-stellar object based on the light profiles of a detection; a `merged' classification exists where an object falls between these two types. A more complete description of the APM sky catalogues can be found at URL {\tt http://www.ast.cam.ac.uk/\verb=~=apmcat/}.

\begin{figure*}
\begin{center}
\setlength{\unitlength}{1cm}
\begin{picture}(10,9)
\put(12.5,-1.5){\includegraphics{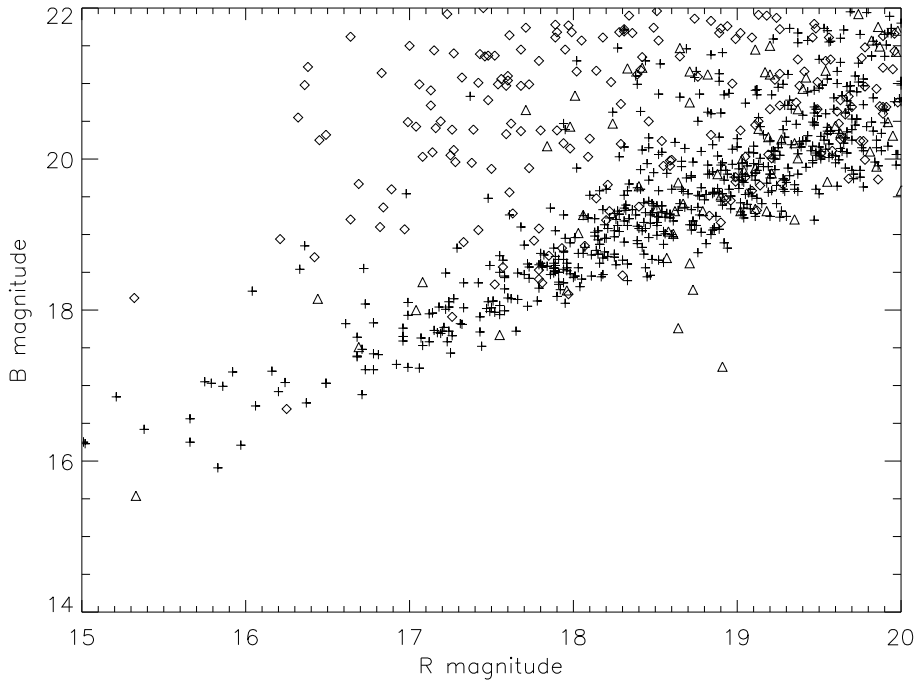}}
\end{picture}
\caption{Plot of R magnitude against B magnitude for a primary source sample where the APM optical classification must be stellar on at least one of the two plates. The diamonds (a) are sources with non-stellar classification on the E (red) plate, the triangles (b) are sources with non-stellar classification on the O (blue) plate and the crosses (c) are sources with a stellar classification on both plates. The range of the plot is restricted to illustrate the main concentration of sources ($\sim$65.5~per~cent of all those with optical identification) - most of those not plotted have a magnitude measured on one plate only. Three separate populations are clearly visible: (a) ($\sim$15.5~per~cent) contains relatively red sources which are probably faint red galaxies; (b) ($\sim$6~per~cent) contains a small population of generally blue sources; (c) ($\sim$78.5~per~cent) are relatively blue stellar sources with a distribution that would be expected for a population of QSOs (for example Newberg et al. 1999). For ARCS, only objects in category (c) were selected (see text).}\label{colour_plots}
\end{center}
\end{figure*}

\subsection{Selection of Primary Sources}
\label{primary_selection}
In order to look for possible cases of multiple imaging we must define a parent sample of ``primary'' sources around which we will search for ``secondary'' images.

With an average beam size of $\sim$5\arcsec, the FIRST survey can confidently identify resolved structure down to an angular scale of $\sim$2\farcs5 (FWHM). For redshifts $z\ga0.2$, this corresponds to a linear scale of $\ga 7$~kpc (for $H_0=65$~km\,s$^{-1}$\,Mpc$^{-1}$). A region of this size will not vary in spectral or luminosity characteristics over periods of hundreds of years. We therefore placed a minimum angular size limit of 2\farcs5 along the deconvolved major axis of our primary sources. We chose the maximum angular size of our primary sources to be 15\arcsec~ to avoid confusion with potentially-lensed secondary images with similar angular separations. 

With the resolution of the FIRST survey many faint radio sources will be resolved into multiple components with angular separations less than 15\arcsec; these components are listed separately in the FIRST catalogue. We therefore constructed two classifications of primary sources, those with a single component and those with two components (see Figure \ref{primary_source}). One could include primary sources with three or more components, but for simplicity we ignored such sources (inclusion of these sources would only increase our sample size by $\sim$10~per~cent). In the case of a single component our maximum primary source size of 15\arcsec refers to the major axis given in the FIRST catalogue, in the case of two components it refers to the angular separation of the components. We defined the central position of a two-component primary source to lie on the line connecting the individual component positions, weighted by the integrated component flux densities (see for example Macklin 1981). The ``fractional centroid distance'' being the fractional distance to the centroid from the brighter component is defined as $\beta=F_B/(F_A+F_B)$ where $F_A$, $F_B$ are the integrated flux densities of the brighter and dimmer components respectively. We can then write the angular distance from the brighter component to the weighted central position as $\theta\times\beta$, where $\theta$ is the angular separation of the two components. Two extreme cases of composite primary sources are illustrated in Figure \ref{composite_source}.

In order to select ``clean'' primary fields and to avoid double counting, we placed a window with a radius of 15\arcsec centred on the central position (defined above) of each primary source. We only included a source in our primary sample if no additional components were within the annulus. This will help prevent more complex (i.e. $3+$ component) sources from infiltrating our sample. We finally imposed the condition that the integrated flux density must be $\ge$35~mJy.

To increase the probability of lensing, it is desirable to select background sources with as high a mean redshift as possible. To this end we selected only those primary radio sources with stellar optical counterparts in the APM catalogue on the O and/or E plate. By choosing stellar objects with associated radio emission we expect to select a high proportion of quasars, and hence to maximise the mean redshift (see section \ref{redshift}). Radio sources with non-stellar i.e. galaxy identifications will tend to have redshifts $\le0.2$ and therefore have a greatly reduced probability of being lensed. Optical identifications with a limiting E magnitude brighter than 20 also make it possible to characterise the redshift distribution of the primary sample with optical spectroscopy using moderate-sized telescopes. In addition, an optical identification on a primary source will provide a filter against selecting classical double radio sources as lens candidates (see section \ref{initial_selection}).

A significant number of faint red galaxies are identified as non-stellar on the E plates, but have a stellar classification on the O plates due to the detection of only the central region of the galaxy. Figure \ref{colour_plots} illustrates the three distinct populations present in a primary source sample when we allowed there to be a non-stellar identification on one of the POSS plates. To simplify the optical identification process for ARCS, an optical object with different classifications on the two plates is not acceptable as a quasar identification. In fact, as Figure \ref{colour_plots} shows, those sources with stellar classifications on one plate and stellar or no classification on the other plate are predominantly blue, as would be expected for quasars (see for example the catalogue derived from four SDSS filters described by Newberg et al. 1999). 

To take into account the positional accuracies of the APM catalogue, a 3\arcsec~difference between the APM and radio positions was tolerated (see section \ref{separation_section}). Radio positions for the single component sources were taken directly from the FIRST catalogue, whereas the composite source positions were as defined above. After applying all these selection criteria, a final parent sample of 1023 primary sources remained (see Table \ref{summary_table}).

\begin{table}
\begin{center}
\caption{Synopsis of the source selection stages of the ARCS project.}\label{summary_table}
\begin{tabular}{lr}
 & Number of sources  \\
FIRST catalogue, 98feb04 & 382\,892  \\
After radio selection (1/2-component) & 7668/5264 \\
After optical selection (1/2-component) & 627/396 \\
Total ARCS primary sample & 1023 \\
Initial lens candidates & 91 \\ 
VLA follow-up candidates (after &  \\
by-eye examination of FIRST maps) & 38 \\
MERLIN follow-up candidates & 2 \\
Number of lens systems & 0 \\
\end{tabular}
\end{center}
\end{table}

\begin{figure}
\begin{center}
\setlength{\unitlength}{1cm}
\begin{picture}(5,6.3)
\put(-2.15,6.8){\includegraphics{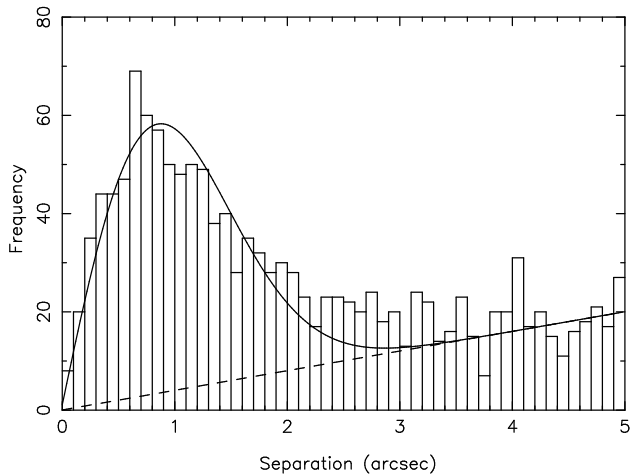}}
\end{picture}
\caption{Histogram of the separation between the optical and the radio position for all primary sources. The dashed line shows the increasing number of random associations with search radius, while the solid line shows the best fit Rayleigh-plus-random distribution, for which $\sigma$=0\farcs86. Whilst our selection criterion allows a maximum optical-radio separation of 3\arcsec, we have extended this plot to 5\arcsec to illustrate the increasing dominance of random associations above 3\arcsec.}\label{separation}
\end{center}
\end{figure}

\subsection{Statistics of the Primary Source Sample}
\subsubsection{Random Radio Associations Posing as Two-component Primaries}
By selecting two-component primary sources from a catalogue with the depth of FIRST we will inevitably introduce a contamination to the primary sample in the form of pairs of sources which have no physical association. We expect this number to be small, but we have made a rough estimate of the number of chance associations of unrelated radio sources posing as two-component sources in the FIRST catalogue. Our selection routine found 5264 two-component primary sources from the radio selection criteria alone i.e. prior to the application of the optical criteria. From the known surface density of FIRST sources the number of random associations was estimated to be $\sim$100. This is a contamination of only $\sim$2~per~cent amongst the two-component sources {\it before} our optical criteria have been applied  (i.e. $<$1~per~cent of {\it all} primaries -- see Table \ref{summary_table}). Within our final primary sample we expect this percentage to have been reduced significantly because randomly associated pairs of sources will be less likely to have stellar counterparts at their centroid position in comparison to a genuine physical pair. Our conclusion is that the level of contamination to the primary sample from the selection of false two-component primary sources will be negligible.

\subsubsection{Random Radio -- Optical Associations}
\label{separation_section}
Since the error in the radio astrometry is negligible relative to the optical astrometric accuracy, and we assume that the optical positional error is purely gaussian, then the separation between the optical position and the true (i.e. radio) position, $r$,  is described by a Rayleigh distribution, viz.
\[
P(r)dr=\frac{r}{\sigma^2}\mbox{exp}\left[-\frac{r^2}{2\sigma^2}\right]dr
\]
where $\sigma$ is both the mean and the standard deviation of the distribution. However, there is a small but significant contamination within our primary sample of radio sources with chance associations with, for example, galactic stellar sources. This modifies the number of optical `identifications' by 
\[
N_{ran}(r)dr=\Sigma_{\star}N_{rad}2\pi rdr
\]
where $\Sigma_{\star}$ is the stellar surface density arcsec$^{-2}$ in the APM catalogue from both POSS plates, $N_{rad}$ is the number of primary sources {\it before} optical identification, and $r$ is the distance from the radio source. Given the fact that $\Sigma_{\star}\sim5\times10^{-4}$~arcsec$^{-2}$ within the FIRST region, the total number of random associations amongst the 1023 primary sources is estimated to be $\sim$180 to a radius of 3\arcsec. Figure \ref{separation} shows a histogram of the angular separation of optical and radio positions in our primary source sample. The distribution is well described by a Rayleigh-plus-random distribution, increasing our confidence that we are selecting extragalactic radio sources with quasar identifications. The best fit value of $\sigma=$0\farcs86 agrees well with the quoted error in the APM positions. However, to select as many primary sources as possible we retained our 3\arcsec~ tolerance for FIRST--APM positional discrepancies. In an analysis of the lensing frequency within the sample, we must modify the primary sample size to $\sim$850 sources to take account of the expected number of random associations.

\subsubsection{Mean Redshift}
\label{redshift}
We searched the NASA/IPAC Extragalactic Database (NED) for additional information on our primary sources. A total of 133 sources were found to have a measured redshift at the FIRST position. We calculated values of $\bar{z}\sim1.09$, $\bar{R}\sim17.5$ and $\bar{B}\sim18.1$ for this sub-sample. The total primary sample is on average fainter with $\bar{R}\sim18.6$ and $\bar{B}\sim19.5$. Whilst the sub-sample obtained from NED will be unquantifiably contaminated by a multitude of selection effects, it is reasonable to assume that the total primary sample will have a higher value for $\bar{z}$. In fact, an initial analysis of spectroscopic data from a representative but limited sub-sample of the primary sample suggests that $\bar{z}\sim1.4$ (Phillips et al., in preparation).

\begin{figure}
\begin{center}
\setlength{\unitlength}{1cm}
\begin{picture}(5,5)
\put(-2.75,-6.3){\includegraphics{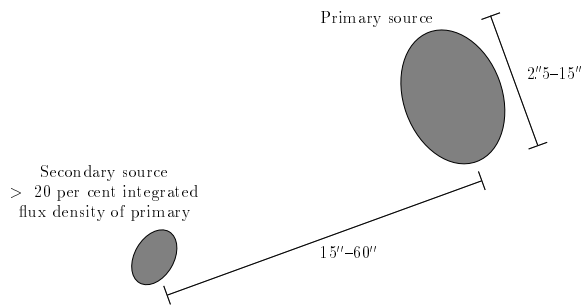}}
\end{picture}
\caption{Illustration of an ARCS lens candidate. Note that while the secondary source in this case is depicted as being smaller than the primary source, in practice no size criteria were applied to the secondary sources.}\label{initial_candidate}
\end{center}
\end{figure}

\subsection{Initial Lens Candidate Selection}
\label{initial_selection}
Having selected the {\it primary} sources the resolution and sensitivity of the FIRST survey allowed us to search for potential multiple imaging events directly. We looked in the catalogue near each primary source for any {\it secondary} sources within an angular separation range of 15\arcsec~and 60\arcsec~(Figure \ref{initial_candidate}); any such secondary sources are putative lensed images of the primary sources. We did not impose any size selection criteria to the secondaries because a secondary lensed image, with a lower magnification than the primary, will not necessarily be resolved by the FIRST beam. We also omitted optical selection criteria; since the secondary will often be weaker, its optical emission could place it below the limit of the POSS/APM catalogue; in addition a stellar object may have varied in magnitude. However, we did impose a maximum flux ratio of 5:1 between the primary and the secondary. This places a lower limit on the integrated flux density of a secondary source in the weakest systems to 7~mJy. If a secondary source is indeed a lensed image of the primary, the conservation of surface brightness means that the secondary will not be resolved out i.e. it will be detectable in FIRST.

In section \ref{primary_selection} we mentioned that a large proportion of extragalactic radio sources will be resolved into multiple components by the FIRST survey. Many such sources consist of two `lobes' of emission, the so-called ``classical doubles''. It is possible that in selecting primary sources with nearby secondaries, we are selecting a high proportion of classical double radio sources as lens candidates. However, the optical identification criterion applied during the primary source selection process provides a strong filter {\it against} these types of radio source posing as lens candidates. This is because the optical counterpart to such an object would be expected to lie {\it between} the lobes rather than on one or both of them as our selection criterion demands.

These selection criteria produced only 91 initial lens candidates i.e. over 90~per~cent of the primary sample could be immediately rejected using the FIRST and POSS/APM information.

\subsection{Further Rejection of Lens Candidates}
\label{arguments}

\begin{figure*}
\begin{center}
\setlength{\unitlength}{1cm}
\begin{picture}(10,7.5)
\put(-4,-2.2){\includegraphics{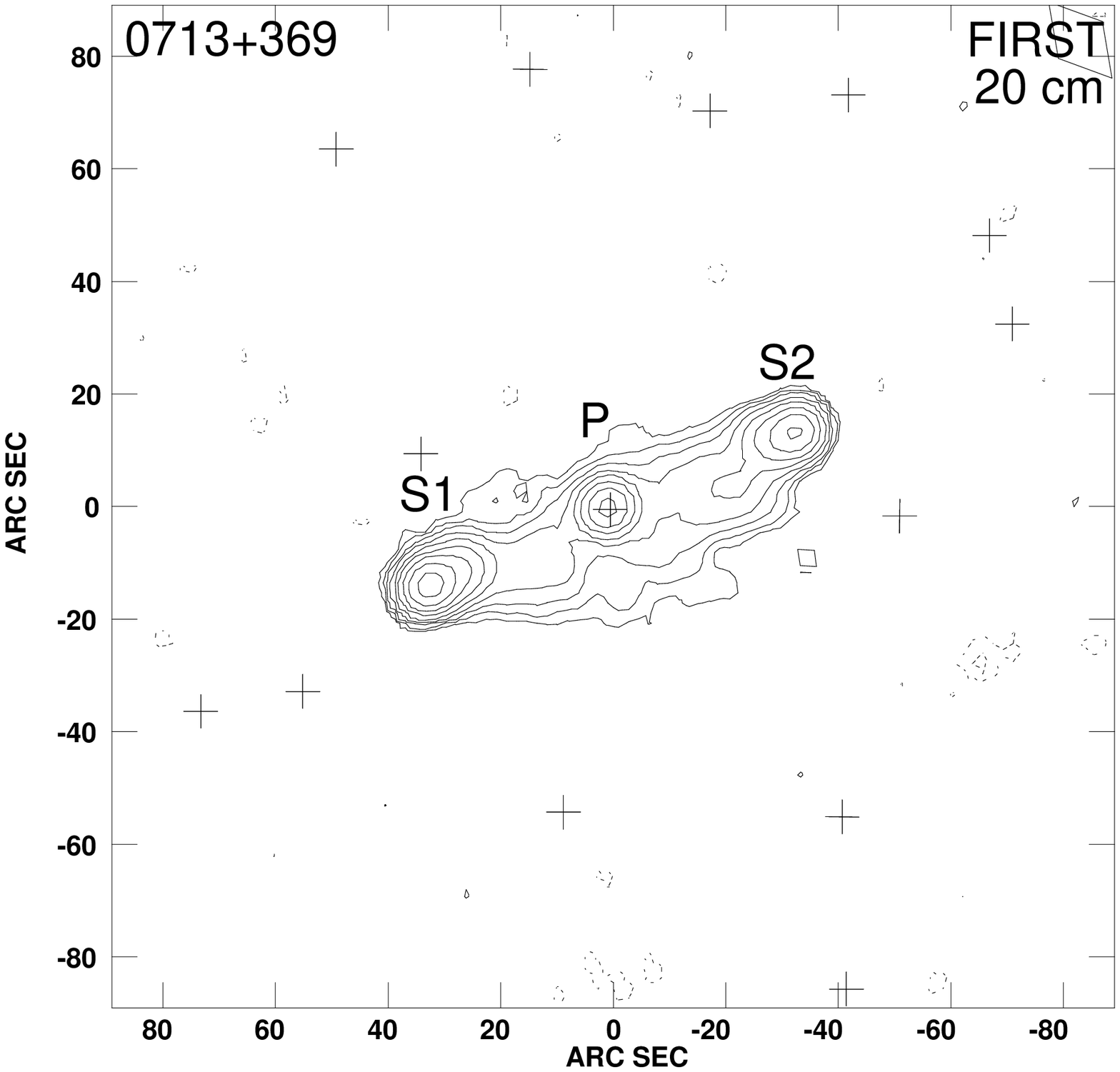}}
\put(4.5,-2.2){\includegraphics{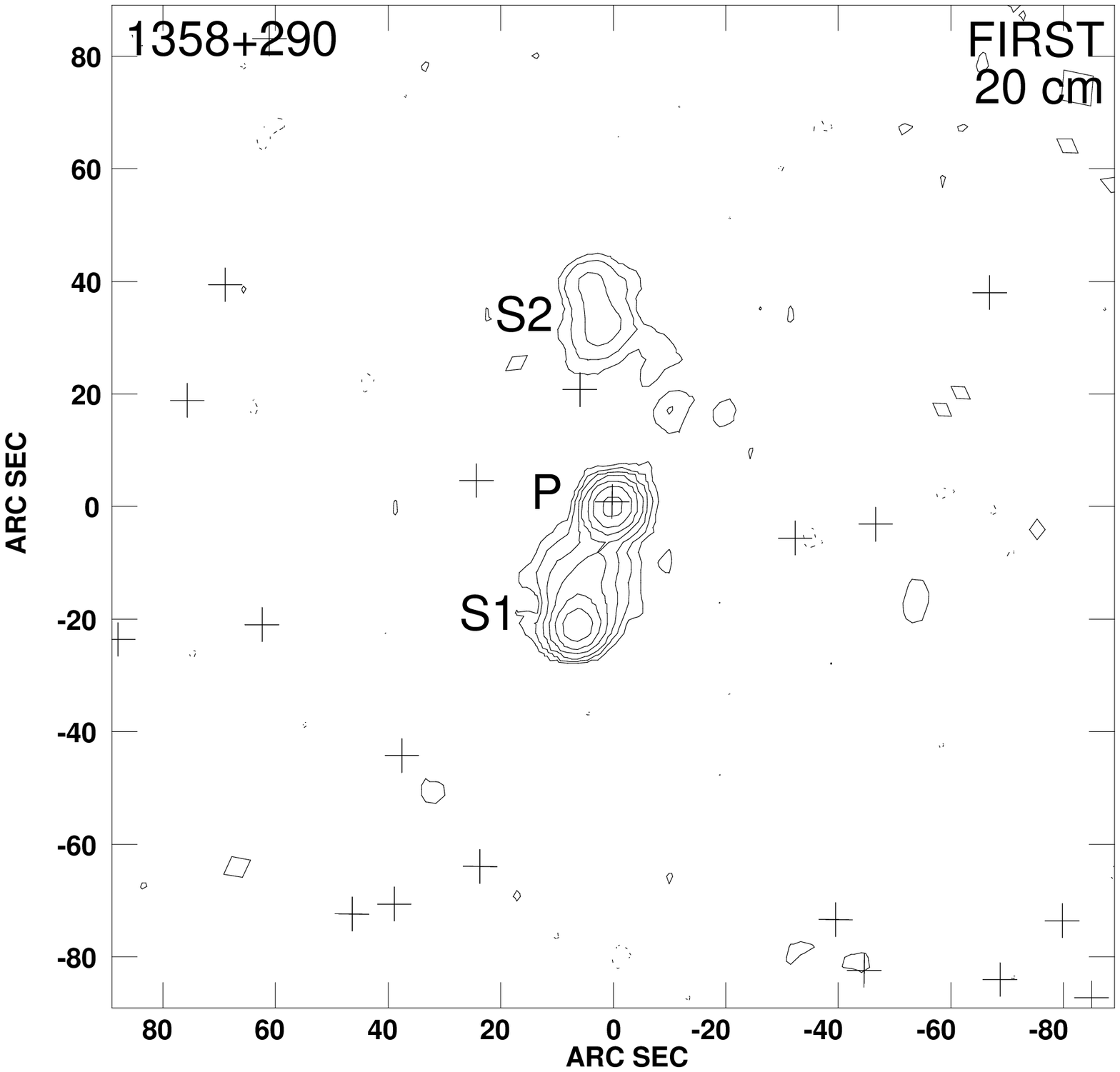}}
\end{picture}
\caption{Examples of FIRST maps from our 91 initial lens candidates. Stellar objects are shown as crosses with a $\pm$3\arcsec~extent, to illustrate the optical identification tolerance; non-stellar objects are shown as diamonds whose dimensions describe the fitted ellipse parameters in the APM catalogue. Radio contours are logarithmic with base 2 starting at 3$\sigma$. Labels P and S denote primary and secondary components respectively. The sources are 0713+369 (left) and 1358+290 (right). The primary `source' is located at the centre of the map in each instance. It is clear that 0713+369 and 1358+290 are single radio sources: 0713+369 has bridge emission between the primary (central) component and the two secondary components are edge brightened away from each other; similarly 1358+290 has a central primary component with a southern secondary component edge brightened away from the primary. The northern secondary component is likely to be a counter-lobe.  Further examples of FIRST maps of ARCS sources are shown in Figures \ref{singles_maps}, \ref{doubles_maps} and \ref{candidates_maps}}\label{FIRST_maps}
\end{center}
\end{figure*}
\begin{figure*}
\begin{center}
\setlength{\unitlength}{1cm}
\begin{picture}(10,22.5)
\put(-4,13){\includegraphics{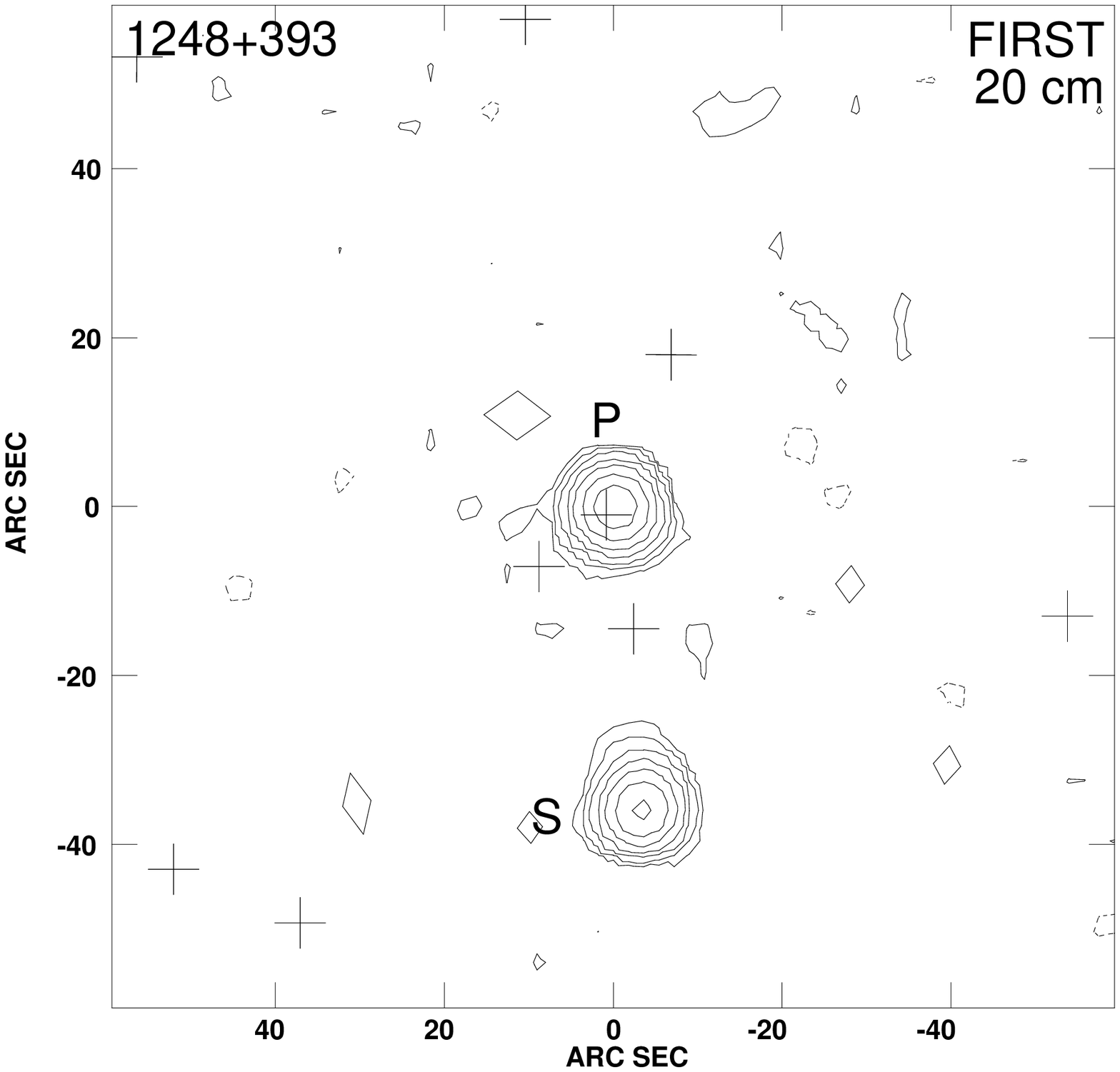}}
\put(4.5,13){\includegraphics{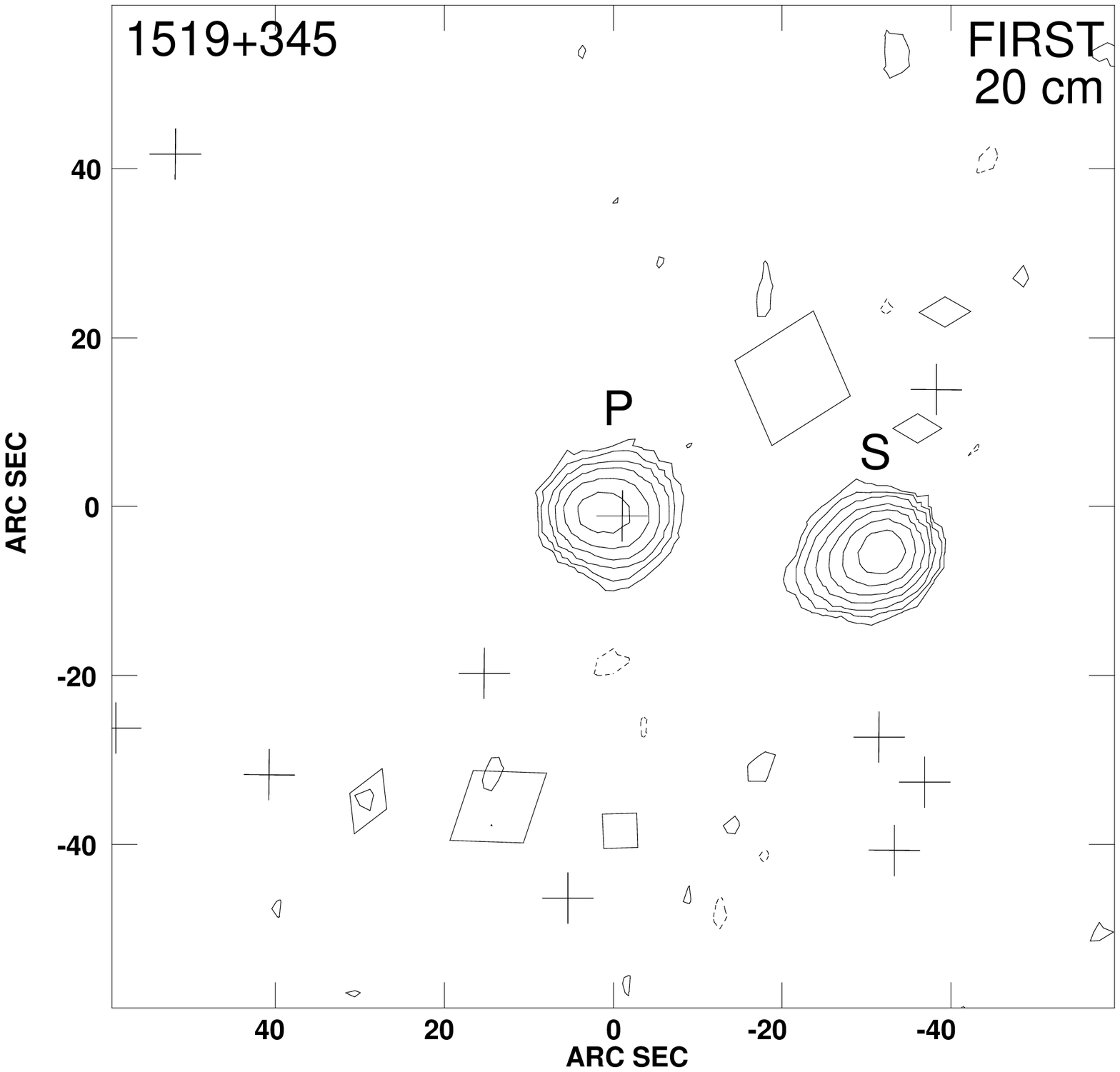}}
\put(-4,5.4){\includegraphics{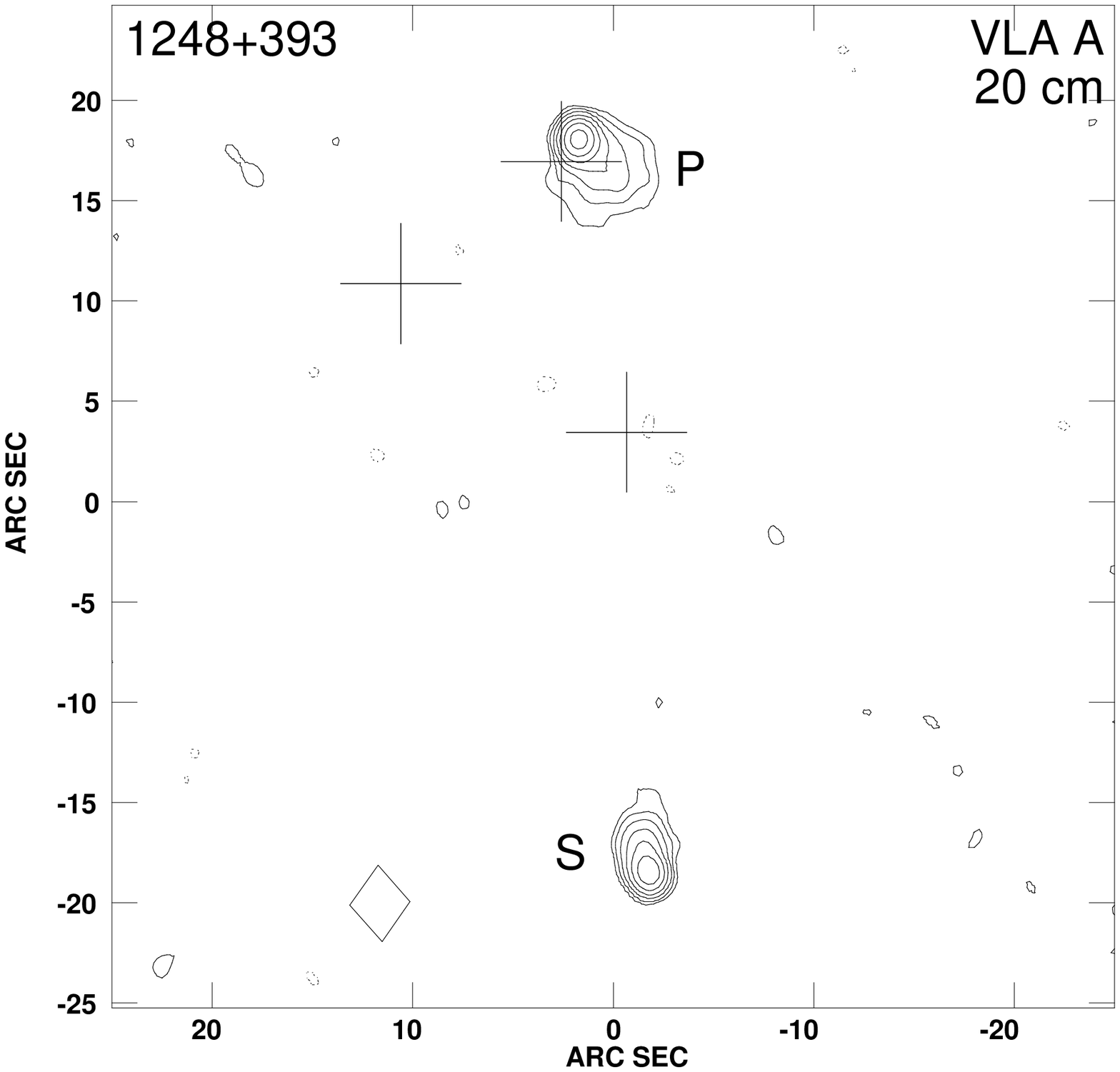}}
\put(4.5,5.4){\includegraphics{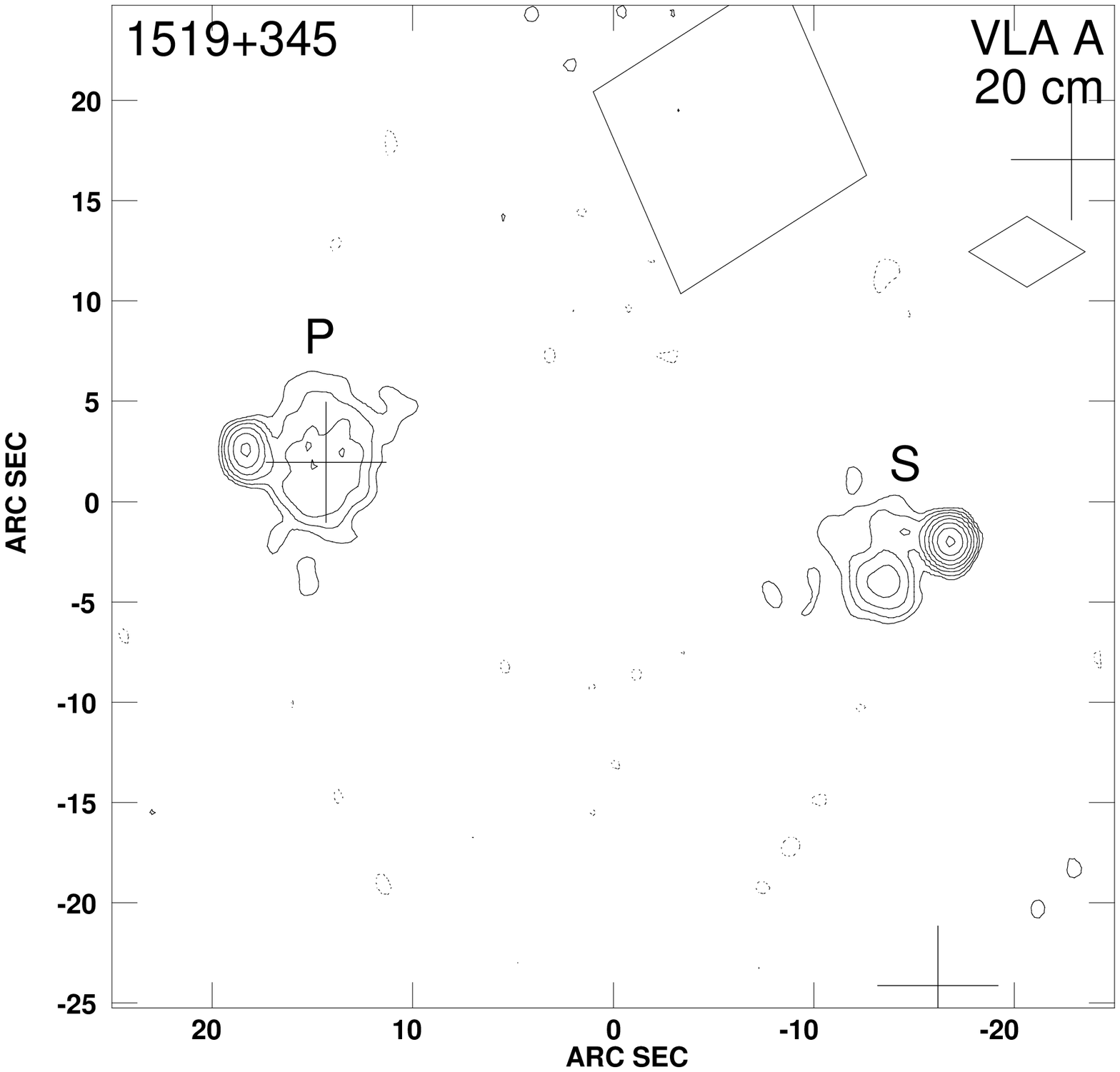}}
\put(-4,-2.2){\includegraphics{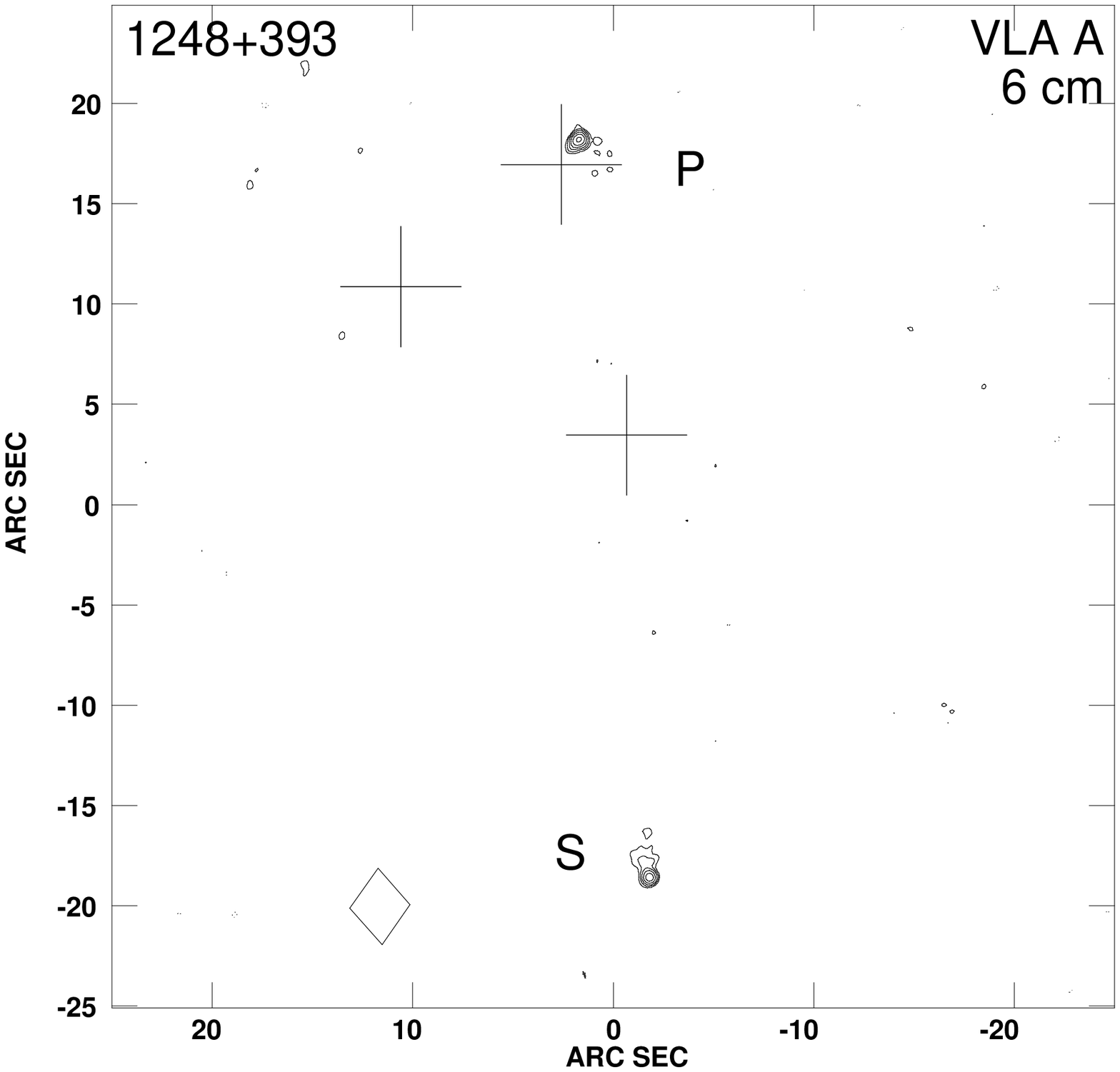}}
\put(4.5,-2.2){\includegraphics{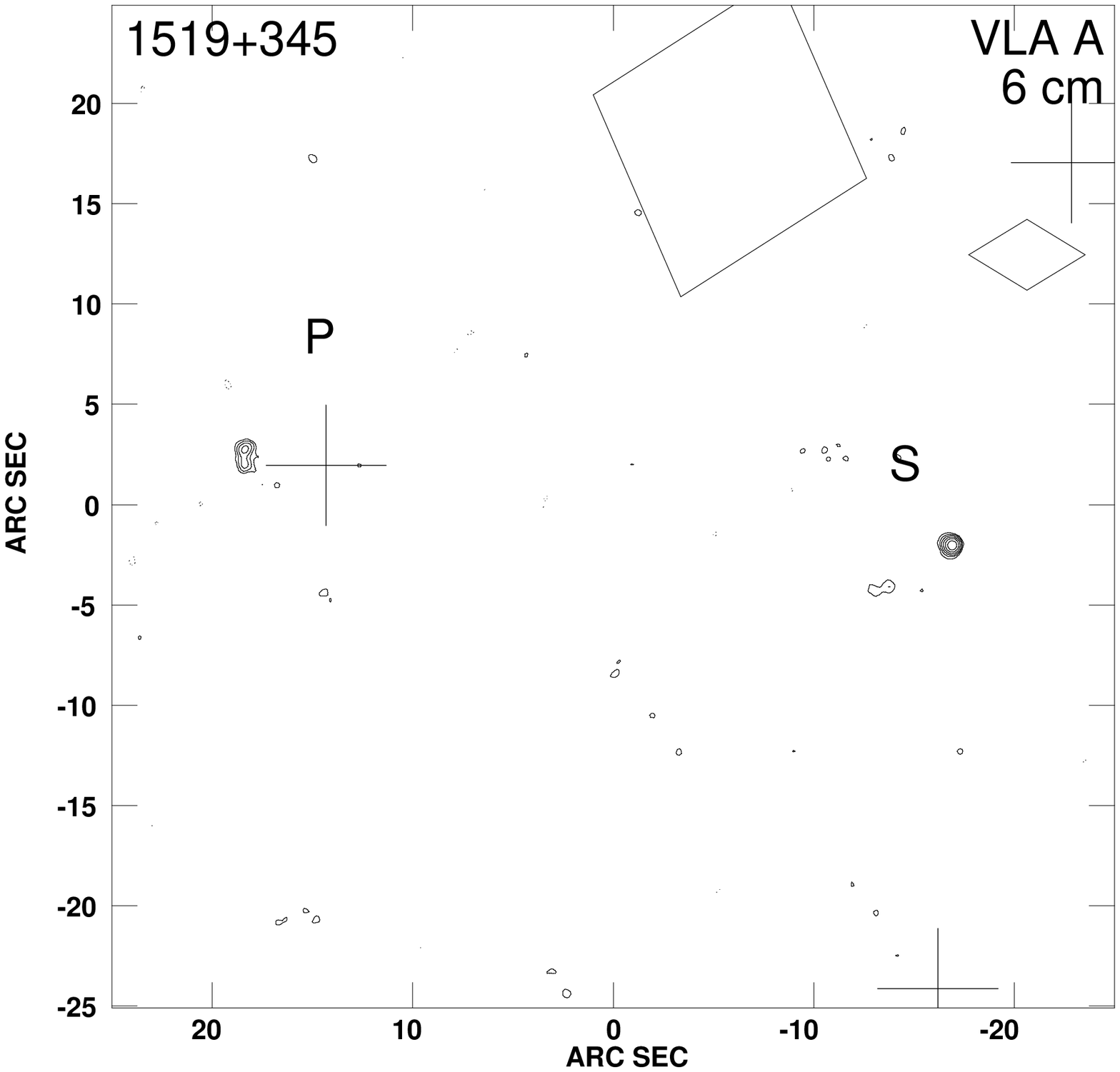}}
\end{picture}
\caption{Examples of single (i.e. intrinsic double) sources from the ARCS VLA follow-up sample: 1248+393 (left) and 1519+345 (right). From top to bottom: FIRST radio map; VLA A configuration 20~cm; VLA A configuration 6~cm. Contours and symbols are as Figure \ref{FIRST_maps}. In both cases it is unlikely that our optical identifications are correct because they are coincident with `lobe' emission. A candidate for the true identification of 1248+393 is located between the primary and secondary radio sources.}\label{singles_maps}
\end{center}
\end{figure*}
\begin{figure*}
\begin{center}
\setlength{\unitlength}{1cm}
\begin{picture}(10,22.5)
\put(-4,13){\includegraphics{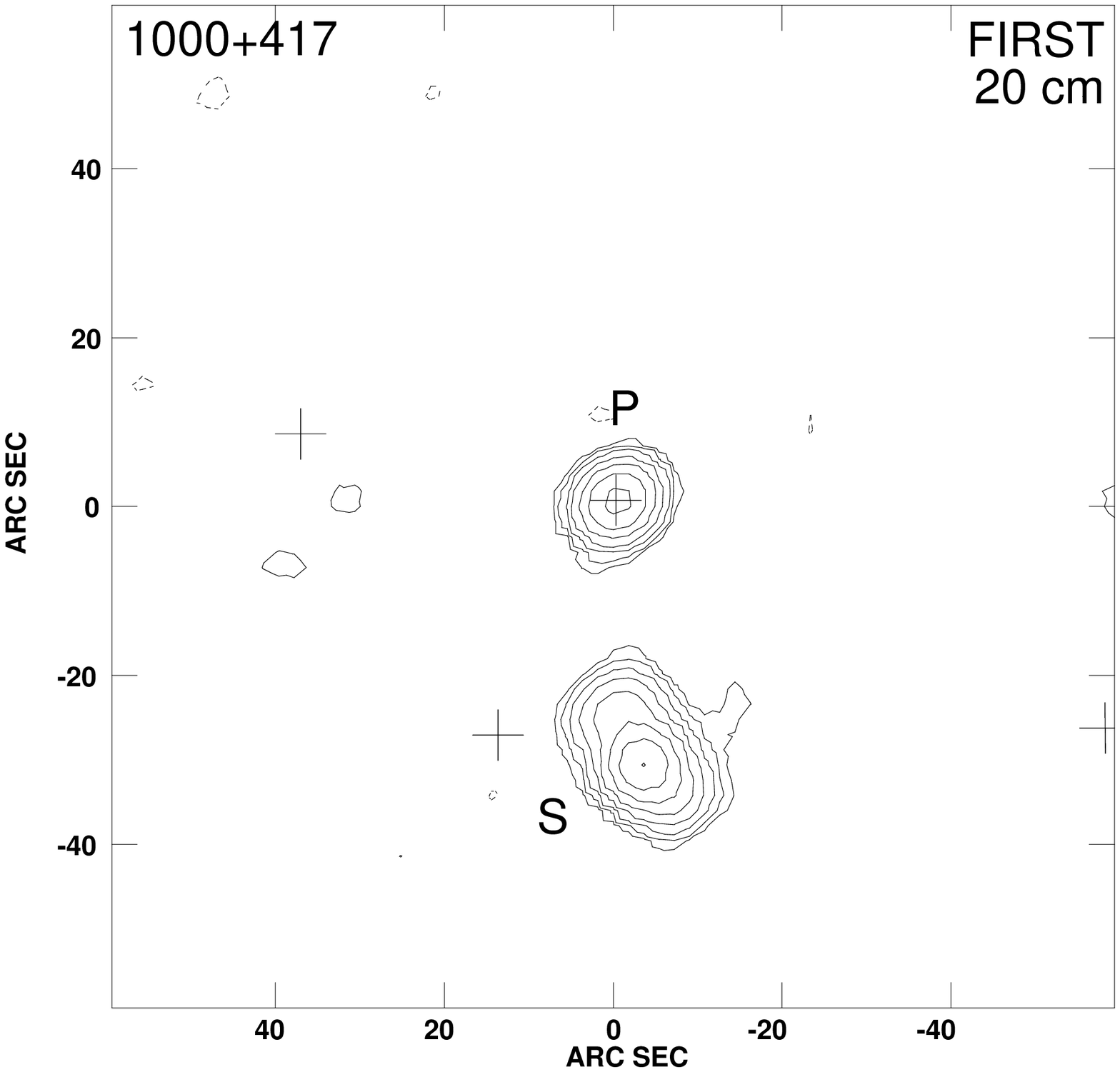}}
\put(4.5,13){\includegraphics{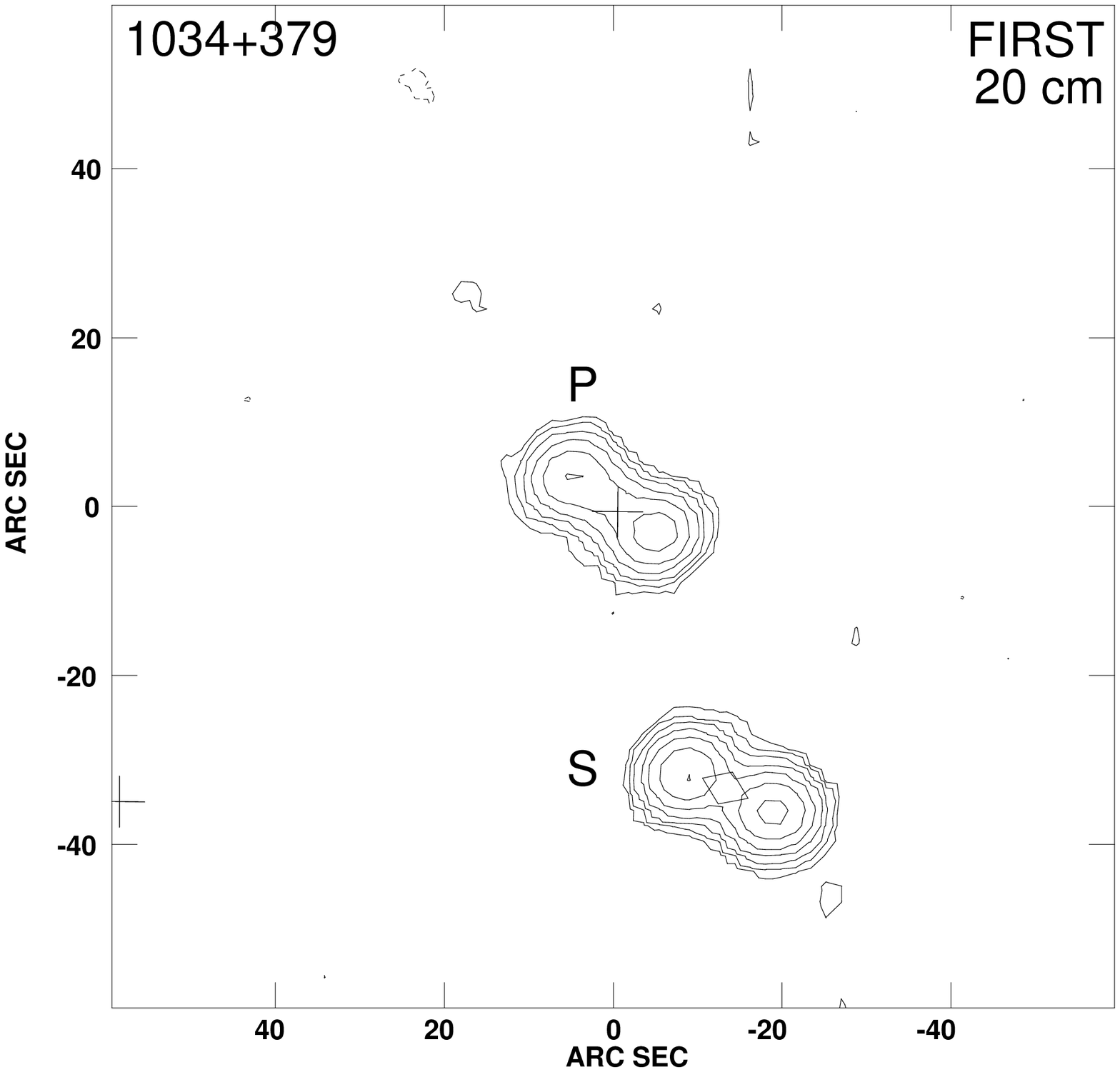}}
\put(-4,5.4){\includegraphics{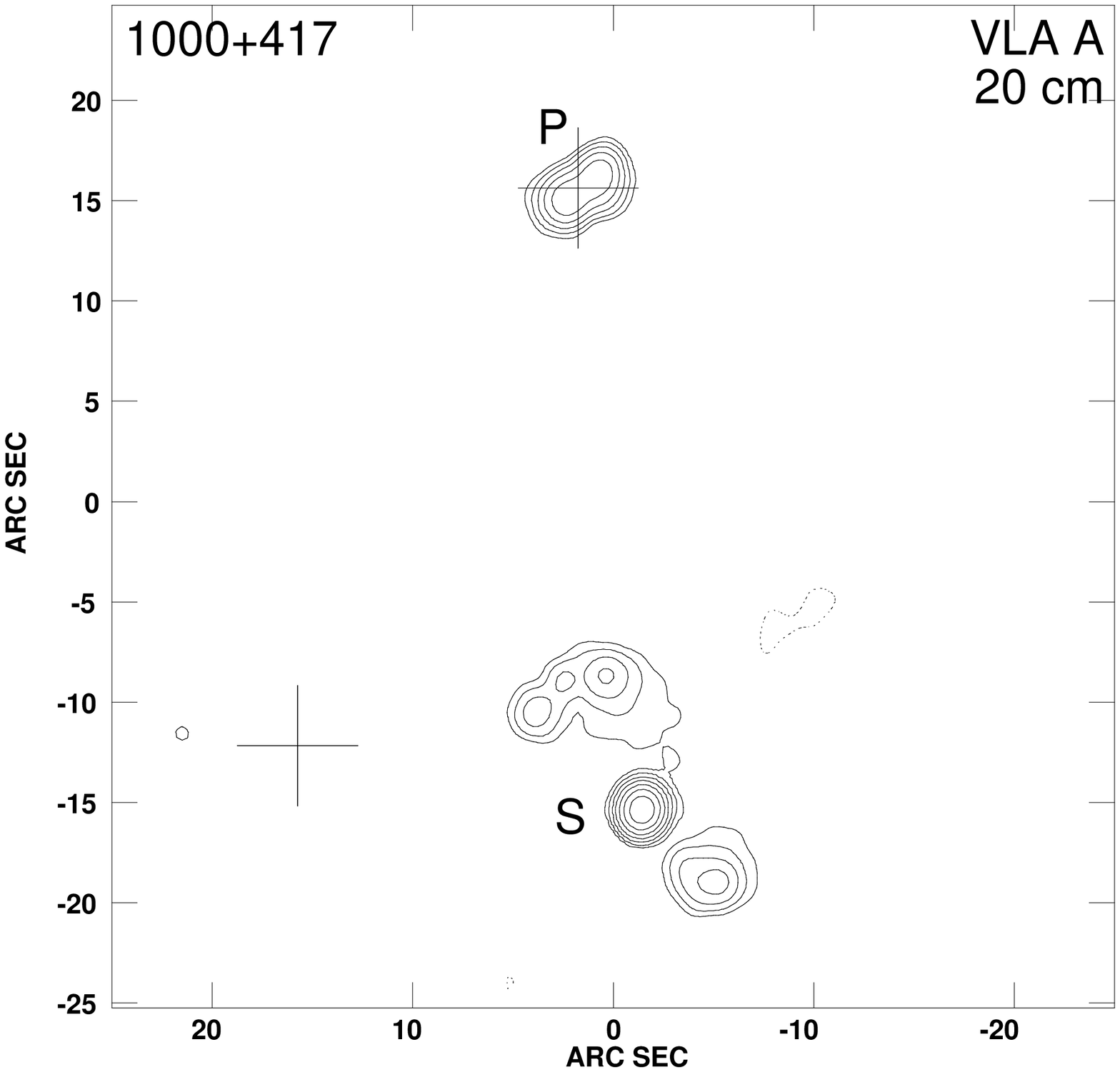}}
\put(4.5,5.4){\includegraphics{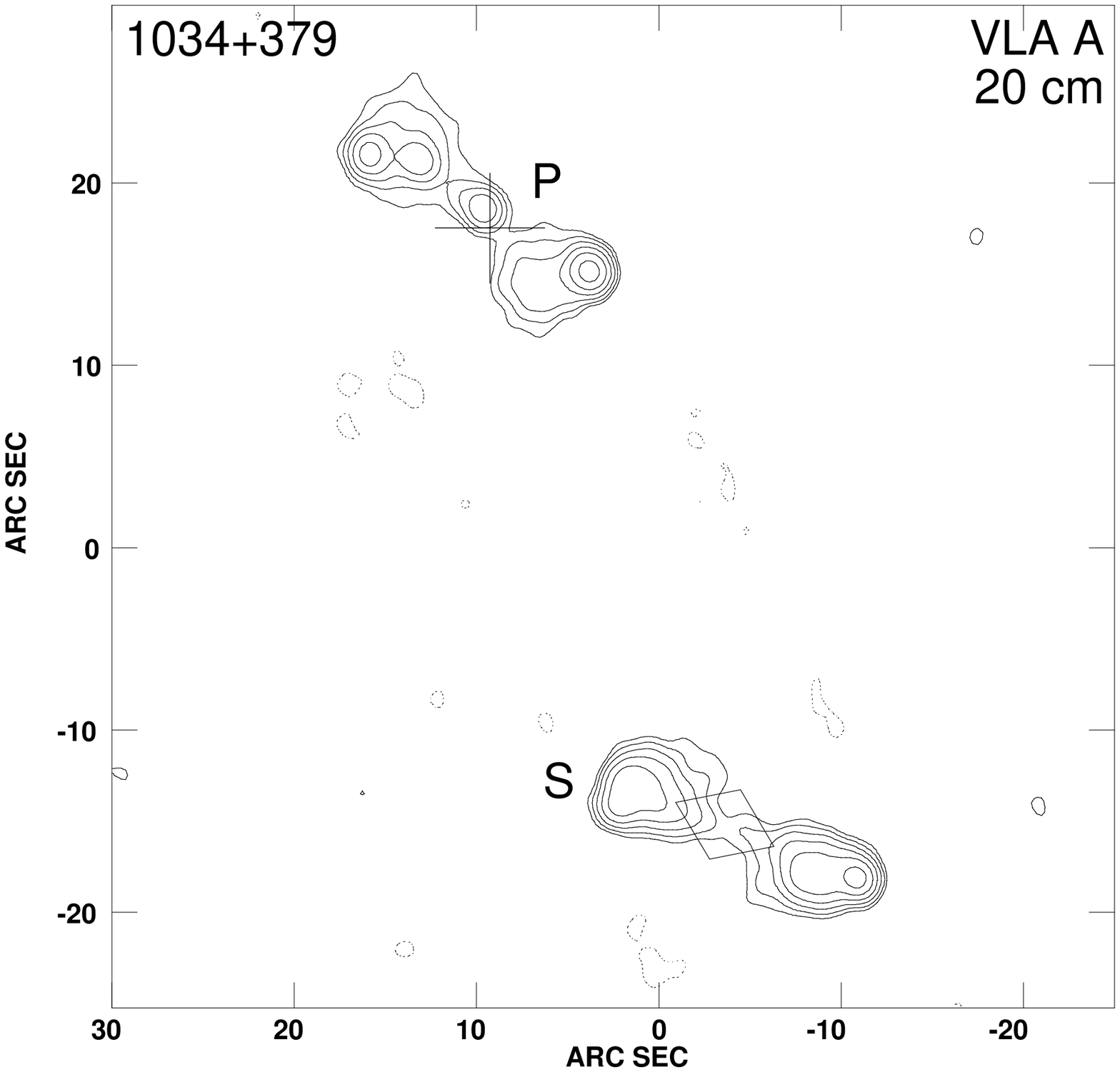}}
\put(-4,-2.2){\includegraphics{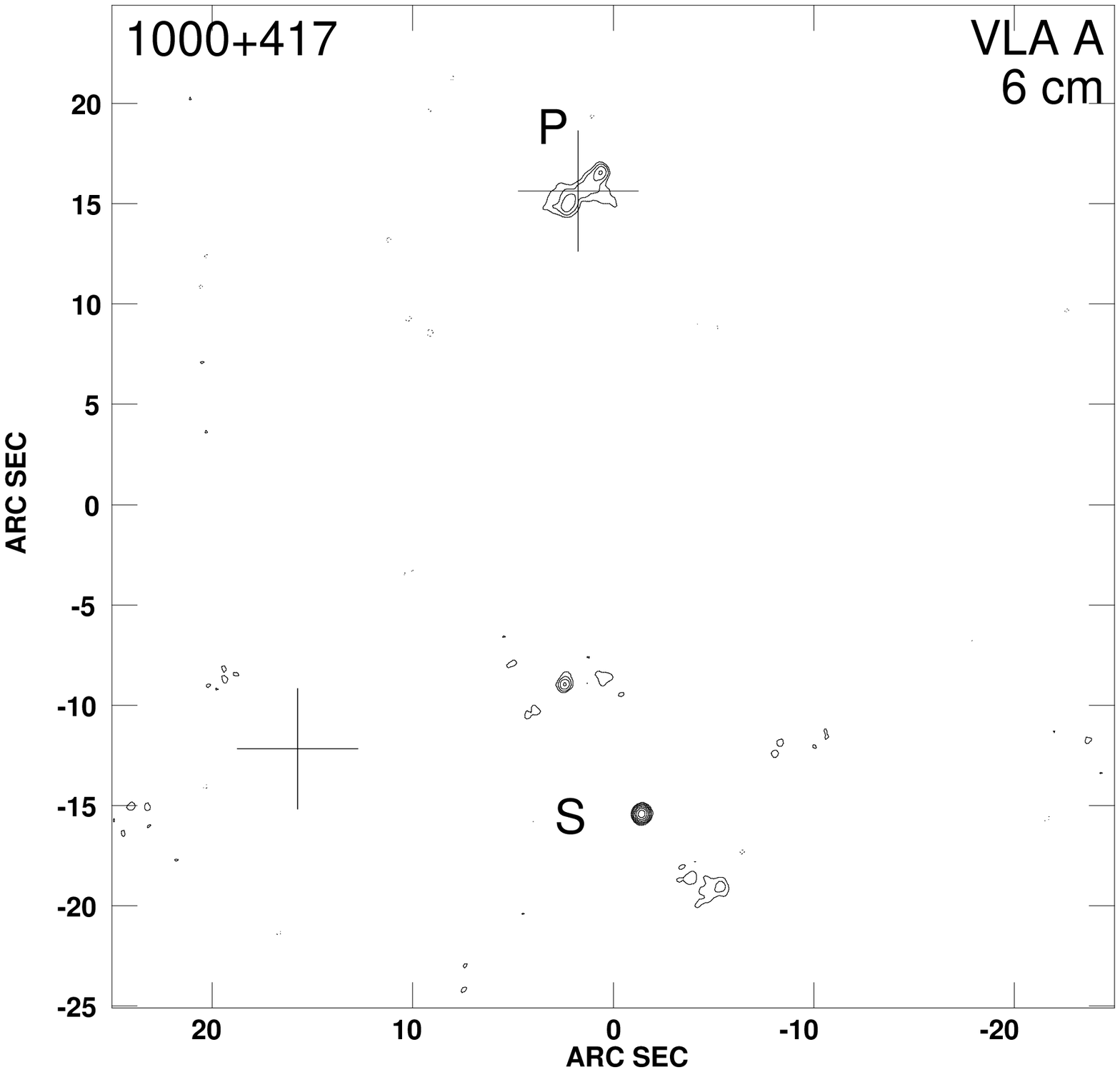}}
\put(4.5,-2.2){\includegraphics{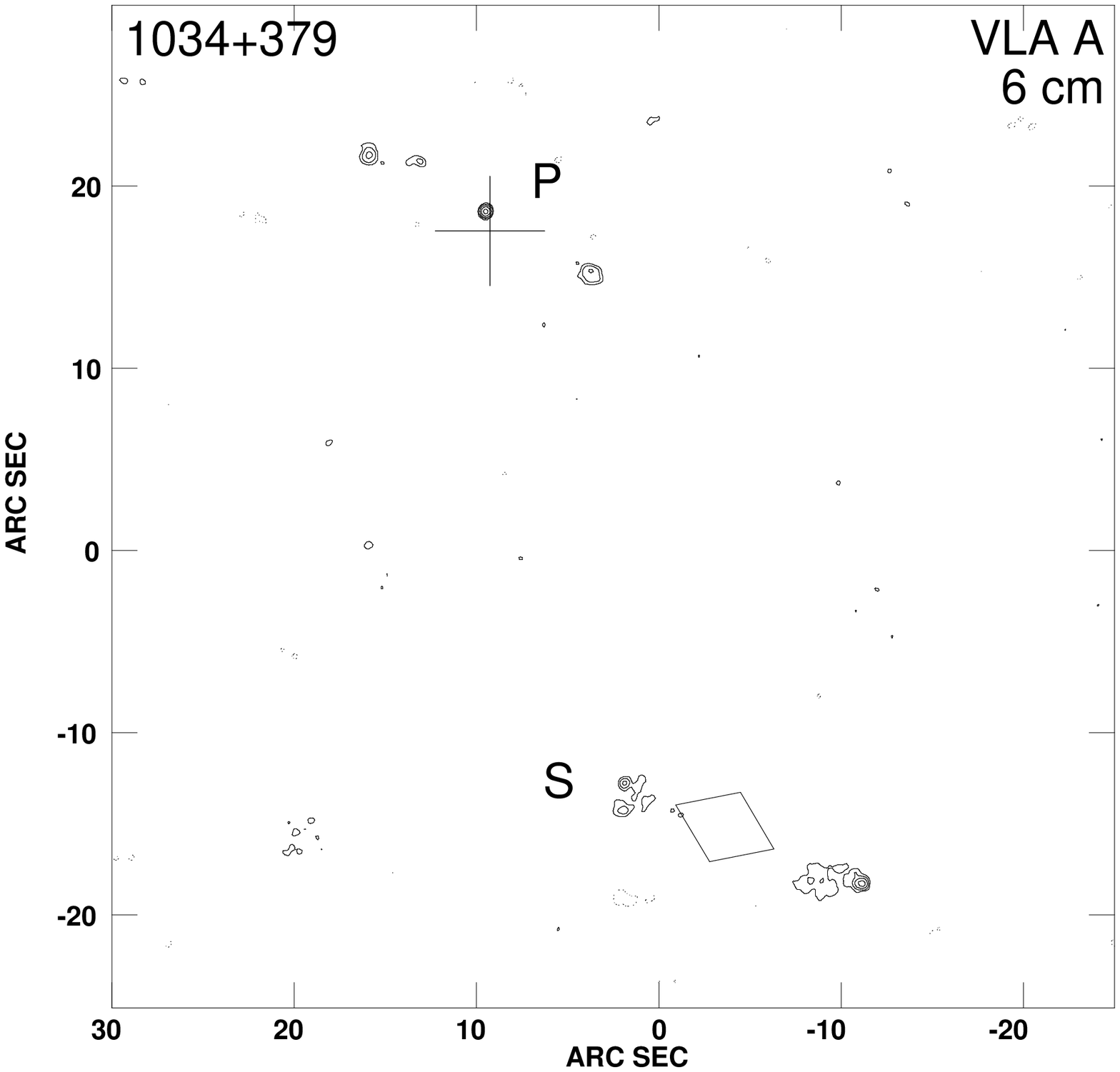}}
\end{picture}
\caption{Examples of pairs of unrelated sources from the ARCS VLA follow-up sample: 1000+417 (left) and 1034+379 (right). From top to bottom: FIRST radio map; VLA A configuration 20~cm; VLA A configuration 6~cm. Contours and symbols are as Figure \ref{FIRST_maps}. In both examples the position of the optical identification of the primary source is consistent with the radio morphology. In addition 1034+379 exhibits a non-stellar optical counterpart between the two lobes of the secondary source.}\label{doubles_maps}
\end{center}
\end{figure*}
\begin{figure*}
\begin{center}
\setlength{\unitlength}{1cm}
\begin{picture}(10,22.5)
\put(-4,13){\includegraphics{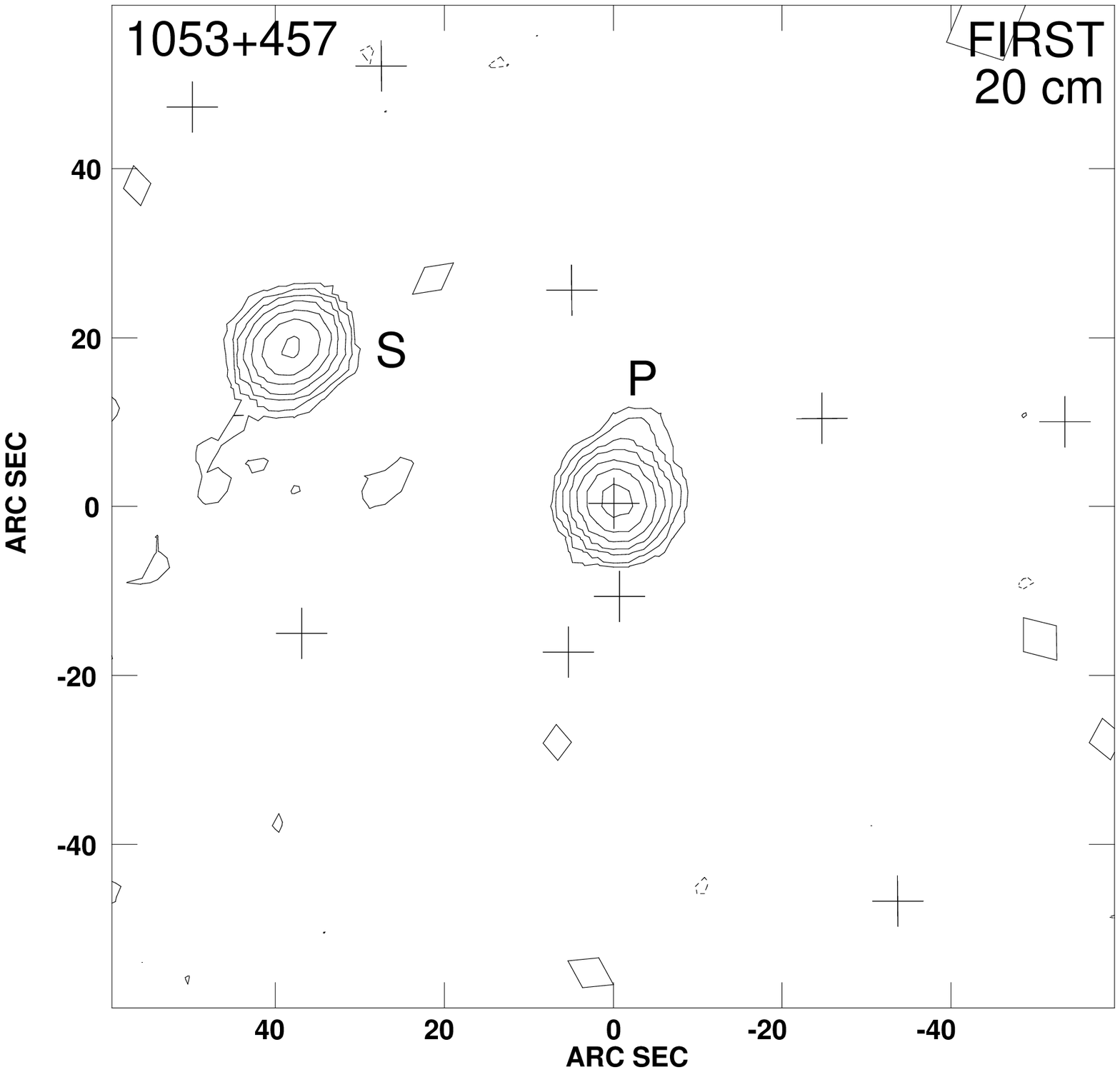}}
\put(4.5,13){\includegraphics{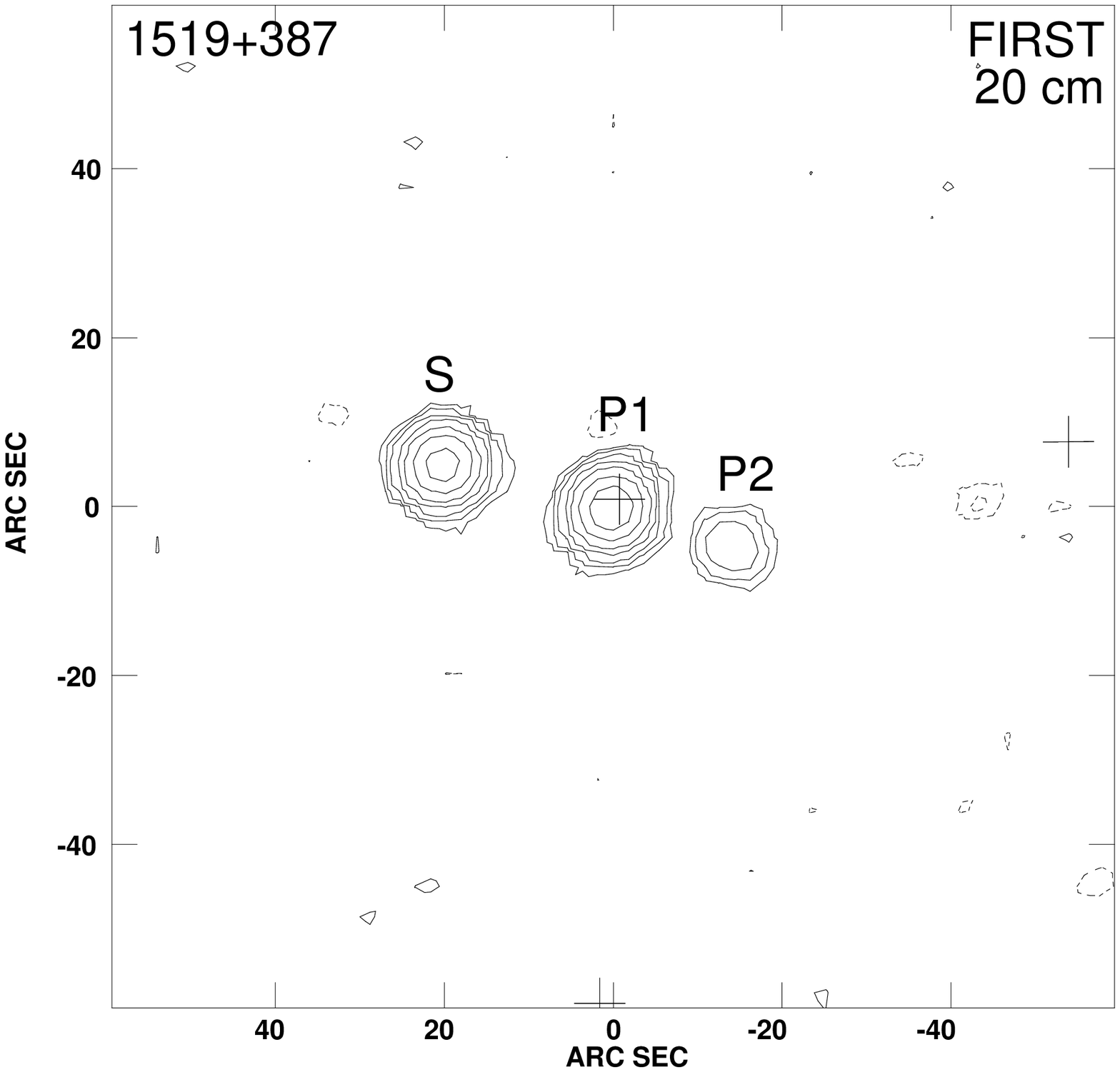}}
\put(-4,5.4){\includegraphics{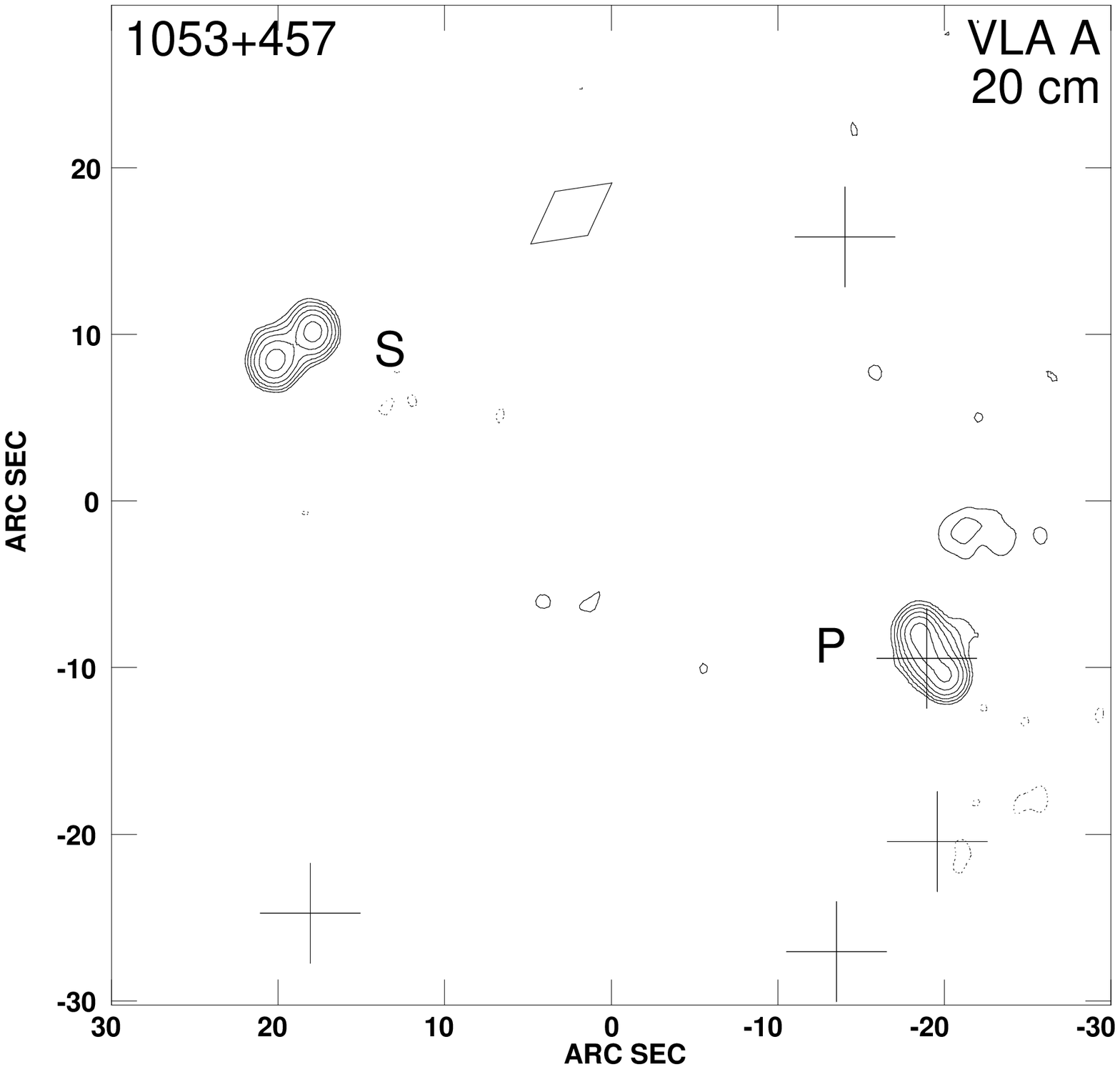}}
\put(4.5,5.4){\includegraphics{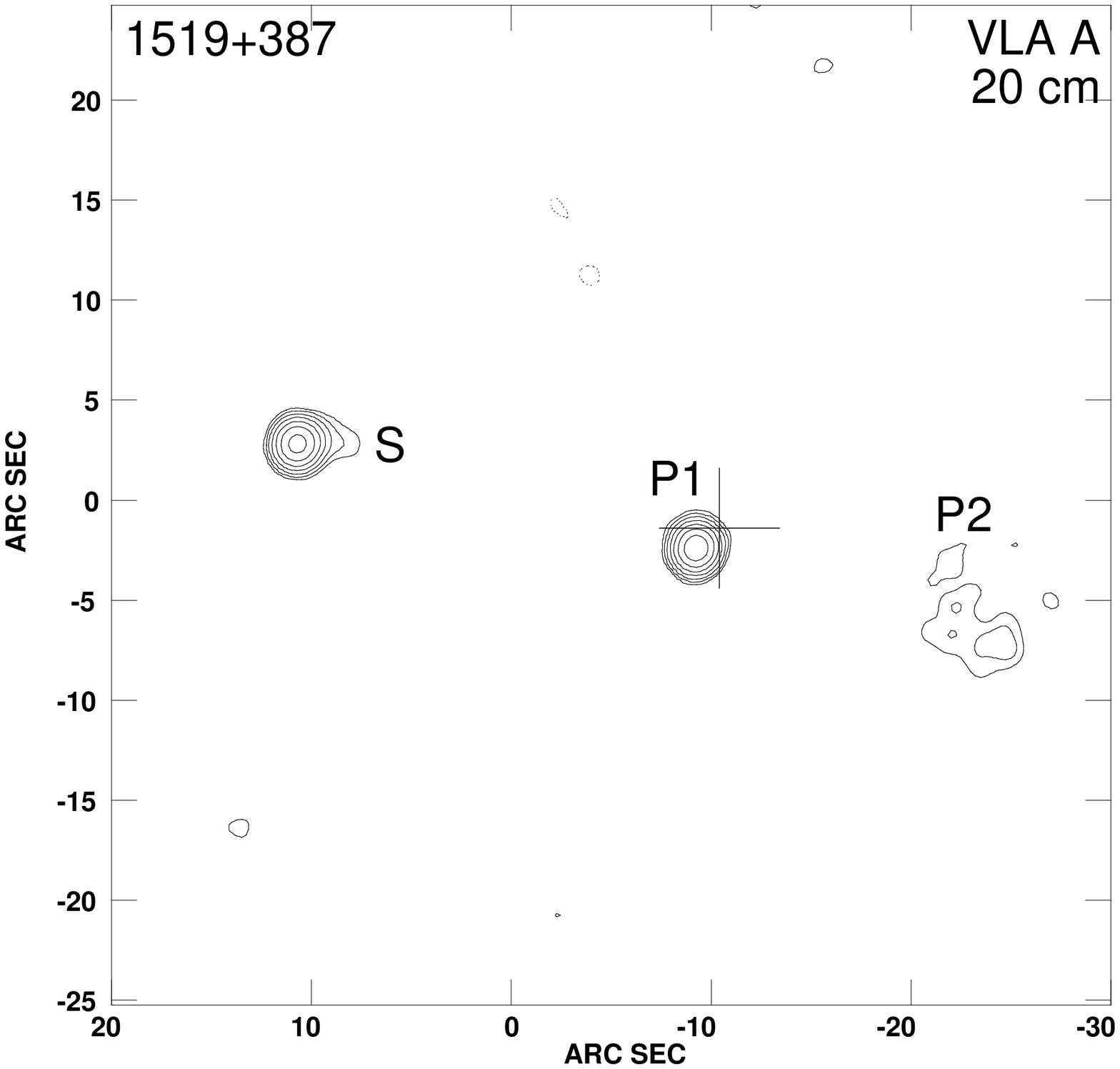}}
\put(-4,-2.2){\includegraphics{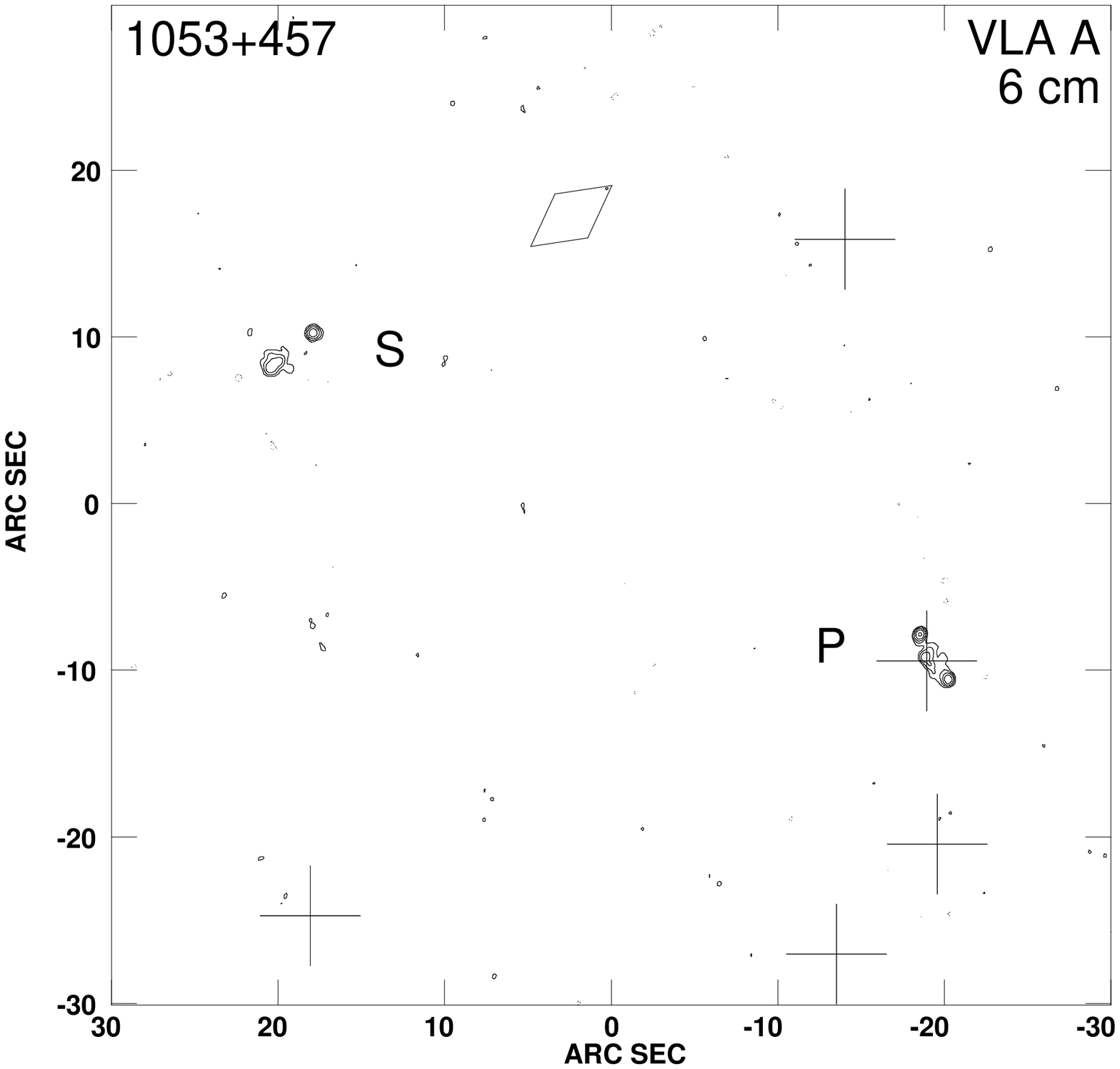}}
\put(4.5,-2.2){\includegraphics{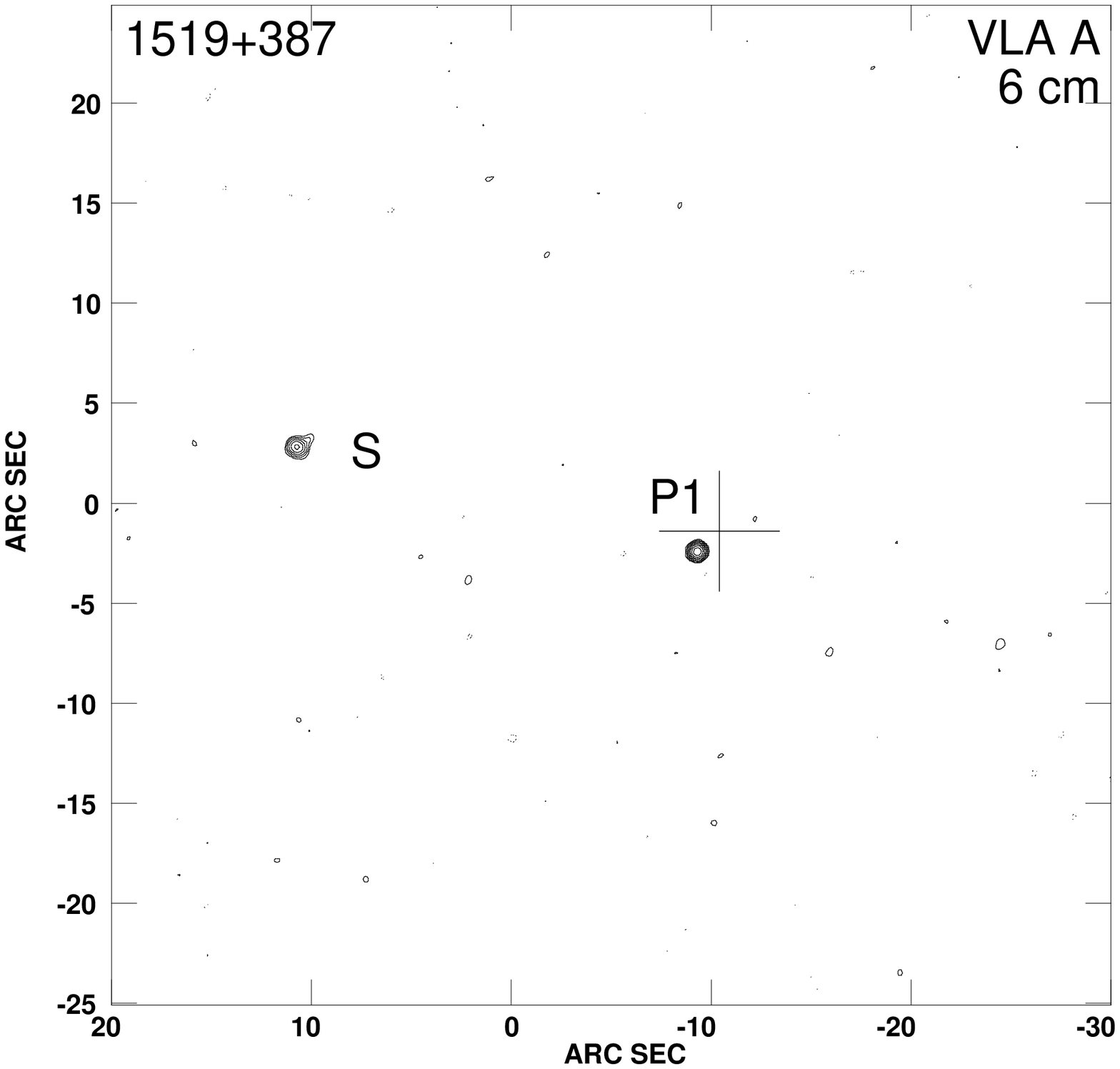}}
\end{picture}
\caption{The two remaining lens candidates from the ARCS VLA follow-up sample: 1053+457 (left) and 1519+387 (right). From top to bottom: FIRST radio map; VLA A configuration 20~cm; VLA A configuration 6~cm. Contours and symbols are as Figure \ref{FIRST_maps}. In both examples the position of the optical identification of the primary source is consistent with the radio morphology.}\label{candidates_maps}
\end{center}
\end{figure*}
\begin{figure*}
\begin{center}
\setlength{\unitlength}{1cm}
\begin{picture}(10,14.7)
\put(-4,5.4){\includegraphics{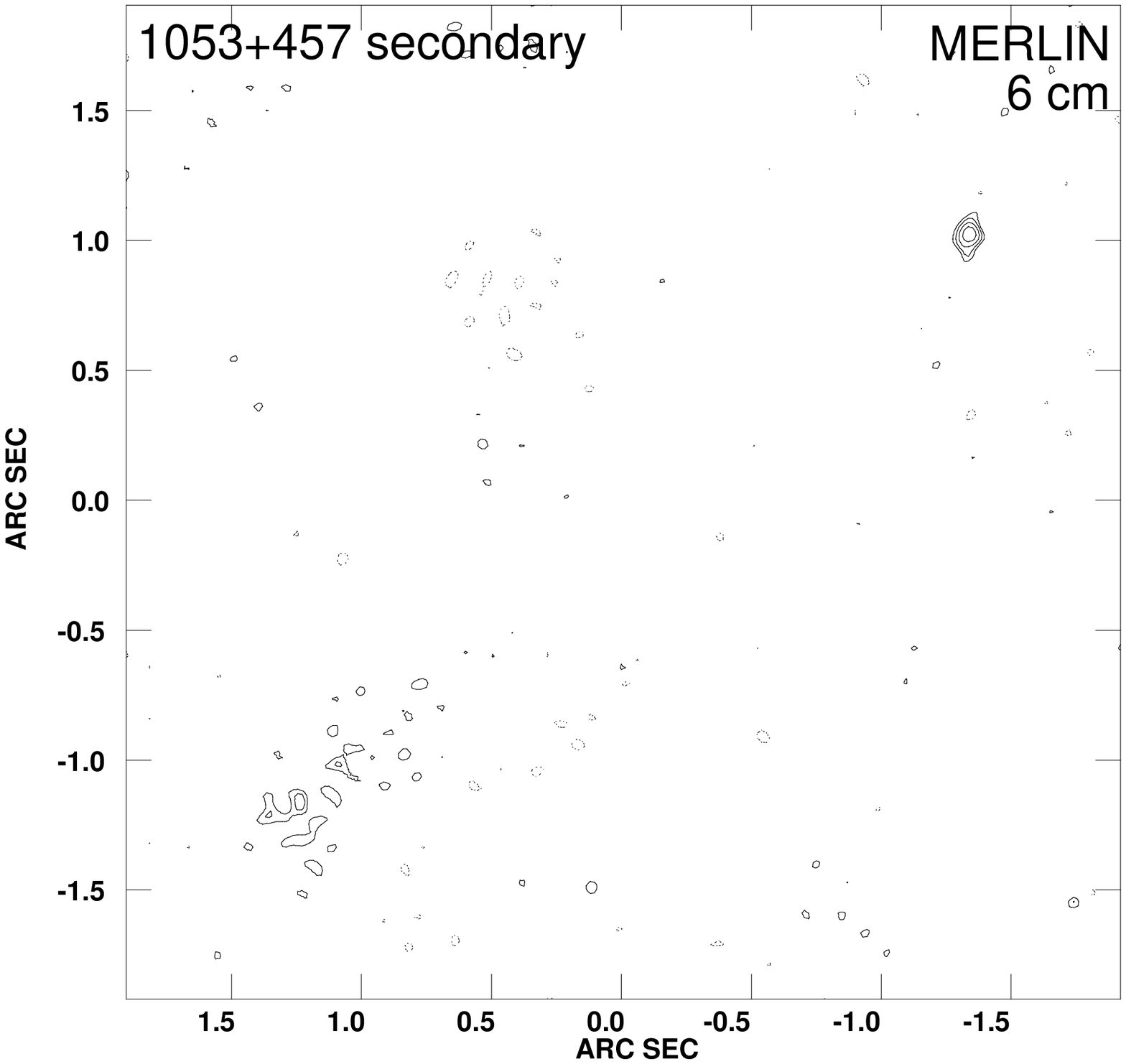}}
\put(4.5,5.4){\includegraphics{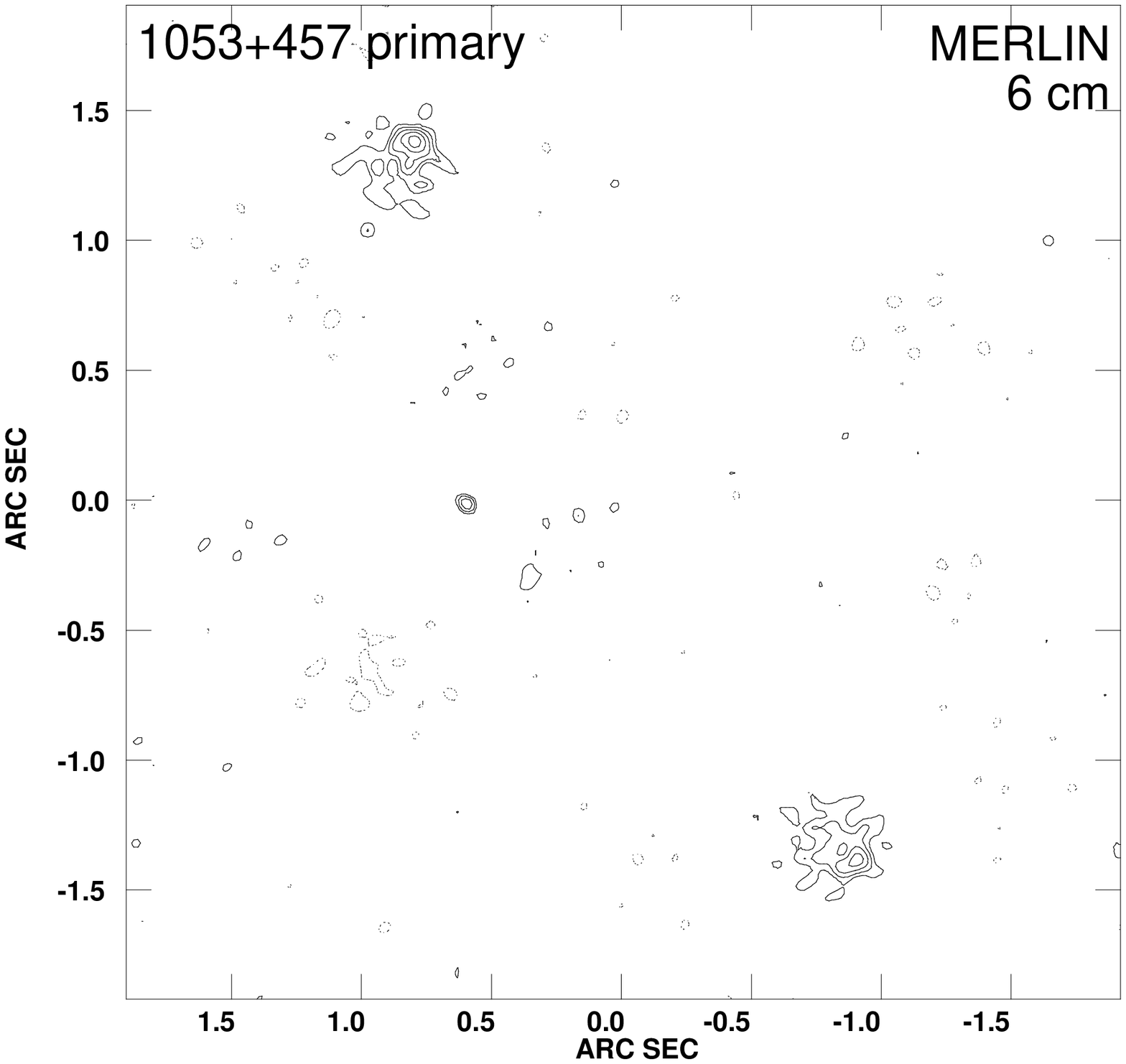}}
\put(-4,-2.2){\includegraphics{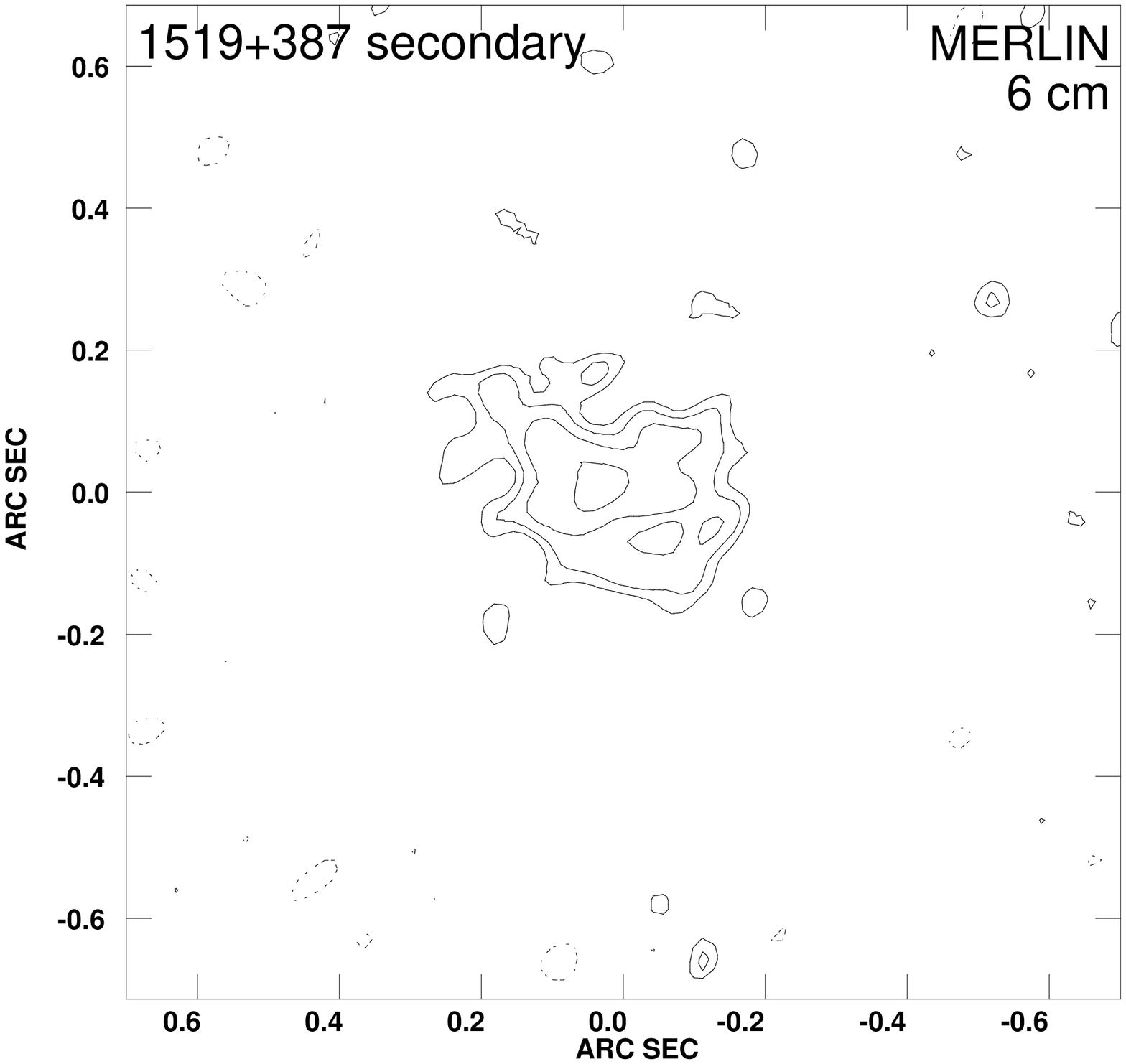}}
\put(4.5,-2.2){\includegraphics{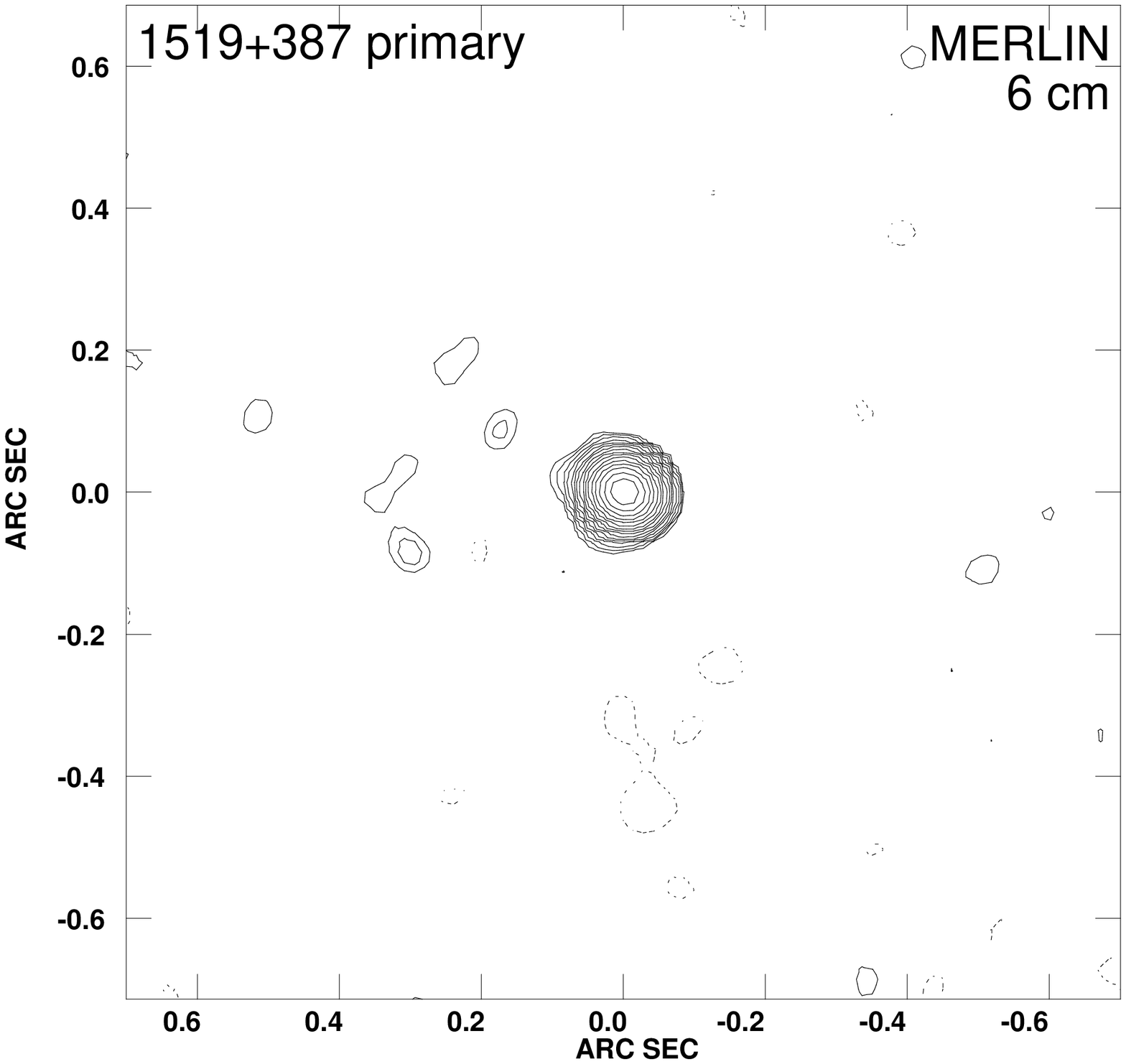}}
\end{picture}
\caption{High resolution 6~cm MERLIN maps of the final lens candidates 1053+457 (top) and 1519+387 (bottom). Individual maps of the secondary (left) and primary sources (right) are shown. In the case of the composite primary source of 1519+387, only the unresolved component is shown. Contours are logarithmic with base $\sqrt{2}$ starting at 3$\sigma$.}\label{MERLIN_maps}
\end{center}
\end{figure*}

Up until this point the lens candidate selection process was conducted automatically. We then went on to examine the FIRST and APM data individually by eye.

The FIRST radio images of the 91 initial lens candidates along with APM information for each field were combined using the AIPS (Astronomical Image Processing Software) package. Examples of typical fields are shown in Figure \ref{FIRST_maps} and the top of Figures \ref{singles_maps}, \ref{doubles_maps} and \ref{candidates_maps}. It is immediately clear that we can reject some more candidates by eye. The arguments for rejecting a system as a lens candidate can be broken down into two categories: (1) those where the primary and secondary sources are clearly components of a single larger source; and (2) those where lensing arguments can be invoked.
\newline
\newline
(1) {\it Single source arguments:}\newline
The single source rejection arguments rely on the nature of known multiple-component radio sources. The morphologies of such sources can include specific features which would not be expected if the components were lensed images of each other. Examples are: extended emission bridging the region between the primary and secondary source, and `outer' edges of a primary-secondary pair being brighter than the `inner' edges. We also reject lens candidates where we find an optical counterpart between the primary-secondary pair coincident with an additional unresolved radio component; this is the position where we would expect an AGN to be present. Examples of sources rejected as single sources on the basis of the FIRST maps are 0713+369 and 1358+290 shown in Figure \ref{FIRST_maps}.
\newline
\newline
(2) {\it Lens arguments:}\newline
We use the basic fact that the surface brightness of the background source is conserved in each lensed image. However, we can only apply this argument to the extended radio structure. The image flux ratio in the radio and optical regimes can be different due to the the variability arguments laid out in section \ref{initial_selection}.

From the original 91 sources with secondary components, 53 could be rejected following this individual examination leaving only 38 lens candidates which required further observations i.e. less than 4~per~cent of the primary sample.

\section{Radio Follow-up of Candidates}
\label{radio_obs}
In order to test the lens hypothesis for the remaining candidates we made higher resolution observations to study candidate source morphology and surface brightness in more detail.

\subsection{Stage 1 -- VLA Observations}
The sample of 38 lens candidates was observed with the VLA in A configuration at 6~cm and 20~cm in two 12-hour periods on 1 and 7 April 1998. The observing bandwidth was 50~MHz in each of two IF channels for both wavelengths, with observing periods of approximately 8~minutes per source per wavelength. The rms sensitivity was $\sim0.05$ mJy at both wavelengths. The minimum beam sizes obtained were 0\farcs4 at 6~cm and 1\farcs4 at 20~cm when sources were observed near the zenith. The data were calibrated in a standard fashion using the AIPS package. Flux calibration was based on two 2-minute observations of 3C286 assuming a flux density of 7.360~Jy at 6~cm, and 14.554~Jy at 20~cm based on the expression derived by Baars et al. (1977). Polarisation calibration was based on three 1.5-minute observations of 1613+342, which was assumed to be unpolarised to correct for antenna effects, while 3C286 was used for phase offset corrections between the two measured orthogonal polarisation systems. Each source was CLEANed at both 6~cm and 20~cm for Stokes parameter I using the DIFMAP software package (Shepherd 1997) and where appropriate self-calibration was applied. Polarisation and spectral index data reduction was performed with the standard AIPS routines. In these cases a uniform convolving beam size of 1\farcs5 was used at both 6~cm and 20~cm for ease of image comparison. A Gaussian taper was applied in the uv plane to the 6~cm data to 30~per~cent at $200$~k$\lambda$ in order to restrict the uv range to that of the 20~cm data. 

We used similar arguments to those described in section \ref{arguments} to classify each system. The 38 lens candidates fall into three distinct groups: single `classical double' radio sources ($\sim$68.5~per~cent); unrelated small angular separation pairs ($\sim$26.5~per~cent); and 2 remaining lens candidates ($\sim$5~per~cent). Figures \ref{singles_maps},\ref{doubles_maps} and \ref{candidates_maps} show examples from each of these groups. 
\newline
\newline
{\it Examples of single sources} (Figure \ref{singles_maps}):\newline
In both examples, 1248+393 and 1519+345, it can be seen from the A configuration 20~cm data that the primary and secondary components are outer edge-brightened. The 6~cm data show that the spectra of the outer edges are flattening, consistent with these regions being hot-spots embedded within diffuse lobes.
\newline
\newline
{\it Examples of two unrelated double sources} (Figure \ref{doubles_maps}):\newline
The 6~cm A configuration map of example 1000+417 shows a compact double primary (northerly) source with a central optical counterpart. The secondary source consists of a central unresolved flat-spectrum component and extended components to the north-east and south-west, with extended emission bridging the entire source. We can rule this example out as a lens candidate using surface brightness arguments. 

In example 1034+379 the 20~cm A configuration data shows extended emission bridging two components in the primary (northerly) source, while the 6~cm data reveals an unresolved flat-spectrum component at the position of the optical counterpart. The secondary source at 20~cm contains two edge-brightened components with a non-stellar optical counterpart between, in a position that a host galaxy would be expected. This example can be ruled out as a lens candidate based on the dissimilar morphologies of the primary and secondary sources, and the orientations of each source relative to one another.
\newline
\newline
{\it Remaining lens candidates} (Figure \ref{candidates_maps}):\newline
The A configuration map of 1053+457 at 20~cm shows two compact double component sources in a plausible lens configuration. The 6~cm data show that the primary (western) source in fact consists of extended emission bridging two unresolved components, whereas the secondary (eastern) source has a single unresolved component with an extended component to the south-east. Despite the dissimilarities between sources, we cannot unequivocally rule out this system as a lens candidate since unresolved radio components can vary significantly over the time-scale of a time delay. The 1519+387 primary (western) source is a composite consisting of an extended western component and an unresolved eastern component in the A configuration 20~cm data. The secondary source appears largely unresolved in the 6~cm data with a possible detection of extended emission towards the unresolved component of the primary source. While these components could be interpreted as a single source with a central unresolved core at P1 (see Figure \ref{candidates_maps}), high resolution MERLIN data are needed to confirm the extension in the secondary source. At this stage it remains as a lens candidate.

\subsection{Stage 2 -- MERLIN Observations}
High resolution data on the two remaining lens candidates 1053+457 and 1519+387 were taken with MERLIN at 6~cm in two 16-hour long-track observations on 12 and 21 December 1998. The data were in 16$\times$1~MHz bands with an integration time of 8~seconds, giving an rms sensitivity of $\sim0.05$~mJy; the minimum beam size was $\sim$0\farcs04. Flux calibration was based on 25~minute observations of 3C286 assuming a flux density of 7.09~Jy. Initial editing of the data was carried out with the standard MERLIN programs with calibration and mapping again performed in AIPS and DIFMAP.

The unresolved components in the VLA 6~cm data of the primary source of 1053+457 are both clearly resolved with MERLIN (Figure \ref{MERLIN_maps}) and appear to be edge-brightened. Additionally a central unresolved component is revealed between the two extended components and is close to an optical object (see Figure \ref{candidates_maps}). The unresolved component in the secondary source also appears to be slightly resolved while the extended component to the south east is almost resolved out. On the lens hypothesis we would expect the two resolved components in the primary source also to be detected in the secondary source. This is not the case. We detect only one resolved component in the secondary, which is also of lower surface brightness than either of the resolved components in the primary source. We can therefore reject 1053+457 as a lens candidate. The secondary source in 1519+387 is clearly resolved with MERLIN while the primary component remains unresolved. We can categorically reject this source as a lens candidate.

\section{Discussion and Summary}
\label{discussion}
We have completed what is currently the largest unbiased survey for multiple image gravitational lens systems with image separations $>$15\arcsec. Our primary sample of 1023 sources was selected from the FIRST and APM catalogues, and our search for lens candidates within this sample was achieved by looking for additional nearby images within the FIRST catalogue. As a result our follow-up required only 38 candidates be observed with the VLA, and just two of these necessitated higher resolution observations with MERLIN. No cases of gravitational multiple imaging with image separations in the range 15\arcsec to 60\arcsec were found. A synopsis of the numbers of sources at each stage of the selection process is given in Table \ref{summary_table}.

Wambsganss et al (1995) made predictions of ``wide-separation'' lensing rates using $N$-body simulations with SCDM parameters (see also Wambsganss, Cen \& Ostriker, 1998). However, the resolution and scale of the simulations were not adequate to provide a reliable estimate of the number of multiple imaging events that we would expect to see in an unbiased search. In particular, cluster mass density profiles must be simulated more accurately because the shape of the profile has a significant influence on a cluster's ability to produce multiple images. In order to draw any significant cosmological conclusions from our result more realistic predictions of the number of wide-separation multiple imaging events in different cosmological scenarios need to be made. The results of such simulations are also necessary if we are to determine the sample size for any future expansion of the ARCS project. 

To examine the practicability of increasing the size of the ARCS sample we have calculated the maximum size of a sample which can be selected from currently available FIRST data and the pending SDSS data. As described in section \ref{FIRST} FIRST has been designed to coincide with the SDSS survey region. It is estimated that $\sim$50~per~cent of all sources in FIRST will be identifiable on SDSS to $m(v)\sim$24 down to the 1~mJy limit of FIRST. Using our radio-only primary source sample of 12\,932 (Table \ref{summary_table}) this could result in a primary sample of $>$6500 if SDSS is used instead of the APM catalogue at the optical identification stage. Taking a more conservative approach by specifying an optical magnitude limit of $m(v)\sim$23 we would still expect to increase our primary sample to $>$3000. It should be noted that these figures are firm lower limits on what could be achieved because our radio-only primary sample has a lower 1.4~GHz flux limit of 35~mJy (c.f. the 1~mJy limit of FIRST). The latest version of the FIRST catalogue dated 99jul21 contains 477\,247 reliable sources over 5450 square degrees of the northern sky (c.f. 382\,892 sources in the 98feb04 version we used). If we also lower the integrated flux density limit of a primary source to 5~mJy, thus giving a minimum secondary flux density at the detection limit of the FIRST survey i.e. 1~mJy, a total radio-only primary sample of 62\,891 sources results. Applying the above optical arguments with a limiting magnitude of $m(v)\sim$23, this should produce a primary sample of $>$15\,000. It would appear that upon completion of SDSS there will be no problem constructing a sample of background sources at least an order of magnitude larger than that presented in this paper.

It is practicable to search such a primary source sample for instances of wide-separation lensing events. With the experience gained in this pilot study, we believe that the efficiency of our search technique could be improved by up to a factor $\sim$2. This is largely based upon our increased confidence in the FIRST survey images for use in rejecting false positives before any further observations of lens candidates are made at higher resolution. Selecting a primary sample of, say, 10\,000 sources, a two-fold improvement in the initial candidate selection would result in only $\sim$200 candidates requiring VLA A configuration follow-up with $\sim$10 requiring further observations with MERLIN. This is equivalent to $\sim$3 days of VLA time and $\sim$3 days of MERLIN time. 

In summary, we have established that it is possible to make a reliable unbiased search for lens systems with image separations in the range 15\arcsec to 60\arcsec. Furthermore it is possible to use existing radio and optical data to significantly reduce the required observing time for such a search. What we currently lack are reliable predictions of the lensing frequency for different cosmologies to compare with our null result.

\section*{Acknowledgments}
We wish to thank the VLA and MERLIN staff for making these observations possible. The VLA is the Very Large Array and is operated by the National Radio Astronomy Observatory which is a facility of the National Science Foundation operated under cooperative agreement by Associated Universities, Inc. MERLIN is the Multi-Element Radio Linked Interferometer Network and is a national facility operated by the University of Manchester on behalf of PPARC. This research has made use of the NASA/IPAC Extragalactic Database (NED) which is operated by the Jet Propulsion Laboratory, California Institute of Technology, under contract with the National Aeronautics and Space Administration. This research was supported in part by the European Commission, TMR Programme, Research Network Contract ERBFMRXCT96-0034 `CERES'. PMP would also like to thank PPARC for the support of a studentship award.

\label{lastpage}


\begin{thebibliography}{99}
\bibitem{b1} Baars J. W. M., Genzel R., Pauliny-Toth I. I. K., 1977, Witzel A., A\&A, 61, 99
\bibitem{b2} Becker R. H., White R. L., Helfand D. J., 1995, ApJ, 450, 559
\bibitem{b3} Browne I. W. A., Patnaik A. R., Wilkinson P. N., Wrobel J. M., 1998, MNRAS, 293, 257
\bibitem{b4} Colley W. N., Tyson J. A., Turner E. L., 1996, ApJ, 461, L83
\bibitem{b5} Irwin M., Maddox S., McMahon R., 1994, Spectrum, 2, 14
\bibitem{b6} Jenkins A., Frenk C. S., Pearce F. R., Thomas P. A., Colberg J. M., White S. D. M., Couchman H. M. P., Peacock J. A., Efstathiou G., Nelson A. H., 1998, ApJ, 499, 20
\bibitem{b7} King L. J., Browne I. W. A., Marlow D. R., Patnaik A. R., Wilkinson P. N., 1999, MNRAS, 307, 2, 225
\bibitem{b8} Lynds R., Petrosian V., 1986, Bull. AAS, 18, 1014
\bibitem{b8b} Macklin J. T., 1981, MNRAS, 196, 967
\bibitem{b9} Maoz D., Bahcall J. N., Schneider D. P., Doxsey R., Bahcall N. A., Lahav O., Yanny B., 1992, ApJ, 394, 51
\bibitem{b10} Maoz D., Rix H., Gal-Yam A., Gould A., 1997, ApJ, 486, 75
\bibitem{b11} Myers S. T., 1996, in Astrophysical Applications of Gravitational Lensing, ed. C. S. Kochanek \& J. N. Hewitt (Dordrecht: Kluwer), 317
\bibitem{b12} Newberg H. J., Richards G. T., Richmond M., Fan X., 1999, ApJS, 123, 2, 377
\bibitem{b13} Patnaik A. R., Browne I. W. A., Wilkinson P. N., Wrobel J. M., 1992, MNRAS, 254, 655
\bibitem{b14} Shepherd M. C., 1997, ADASS VI, A.S.P. Conference Series, 125, 77 
\bibitem{b15} Soucail G., Mellier Y., Fort B., Mathez G., Hammer F., 1987, A\&A, 184, L7
\bibitem{b16} Wambsganss J., Cen R., Ostriker J. P., Turner E. L., 1995, Science, 268, 274
\bibitem{b16b} Wambsganss J., Cen R., Ostriker J. P., 1998, ApJ, 494, 29
\bibitem{b17} Wilkinson P. N., Browne I. W. A., Patnaik A. R., Wrobel J. M., Sorathia B., 1998, MNRAS, 300, 790
\end{thebibliography}
\end{document}